\documentclass[11pt]{article}

\usepackage{scalerel}

\usepackage{amssymb,amsmath,amsthm,mathrsfs,latexsym,amsxtra,graphicx,appendix,epstopdf,feynmf,hyperref,setspace,fix-cm,color,cite,hyperref}
\usepackage{verbatim}
\usepackage[a4paper,left=2cm,right=1cm,top=4cm,bottom=4cm,bindingoffset=5mm]{geometry}
\usepackage{pgfplots}
\pgfplotsset{compat=newest}
\usepackage{mathtools}
\usepackage{tensor}
\usepackage{siunitx}
\usepackage{tikz}
\usetikzlibrary{trees}
\usetikzlibrary{decorations.pathmorphing}
\usetikzlibrary{decorations.markings}
\usetikzlibrary{decorations.markings}
\usepackage{float}
\usepackage{rotating}
\usepackage{simplewick}
\usepackage{cancel}
\usepackage{ marvosym }
\usepgfplotslibrary{fillbetween}
\usetikzlibrary{patterns}

\usepackage{mathtools}

\usepackage{setspace}
\usepackage{lipsum}
 \usepackage{multirow}
\usepackage{fullpage}
\usepackage{amsmath}
\usepackage{amssymb}
\usepackage{setspace}
\usepackage{bbm}
\usepackage{dsfont}
\usepackage{graphics}
\usepackage{subcaption}
\usepackage{longtable}
\usepackage[font=footnotesize,labelfont=bf,justification=centerlast,width=.9\textwidth]{caption}
\usepackage{color}
\usepackage{etoolbox}
 \usepackage{multirow}
\usepackage{booktabs}
\usepackage{array}
\usepackage{hyperref}
\usepackage{cite}
\usepackage[normalem]{ulem}
\hypersetup{
    bookmarks=true,%
    colorlinks,%
    citecolor=blue,%
    filecolor=blue,%
    linkcolor=blue,%
    urlcolor=blue
}
\allowdisplaybreaks
\newcommand{\be}{\begin{equation}}
\newcommand{\ee}{\end{equation}}
\newcommand{\nn}{\nonumber}
\newcommand{\bes}{\begin{equation*}}
\newcommand{\ees}{\end{equation*}}
\newcommand{\bea}{\begin{eqnarray}}
\newcommand{\eea}{\end{eqnarray}}
\newcommand{\beas}{\begin{eqnarray*}}
\newcommand{\eeas}{\end{eqnarray*}}

\newcommand{\dd}{\mathrm{d}}
\newcommand{\bmat}{\begin{bmatrix}}
\newcommand{\emat}{\end{bmatrix}}

\def\le{\left}
\def\ri{\right}

\newcommand{\ZZ}{\mathbb{Z}}

\begin{document}
\numberwithin{equation}{section}
{
\begin{titlepage}

\begin{center}

\hfill \\
\hfill \\
\vskip 0.75in

{\Large \bf Gravitational Wilson Lines in 3D de Sitter}

\vskip 0.4in

{\large Alejandra Castro${}^a$, Philippe Sabella-Garnier${}^{b}$, and  Claire Zukowski${}^a$}\\
\vskip 0.3in

${}^{a}${\it Institute for Theoretical Physics, University of Amsterdam,
Science Park 904, Postbus 94485, 1090 GL Amsterdam, The Netherlands} \vskip .5mm

${}^{b}${\it Lorentz Institute, Leiden University, Niels Bohrweg 2, 2333-CA Leiden, The Netherlands } \vskip .5mm
\texttt{a.castro@uva.nl,  garnier@lorentz.leidenuniv.nl, c.e.zukowski@uva.nl }

\end{center}

\vskip 0.35in

\begin{center} {\bf ABSTRACT } \end{center}
We construct local probes in the static patch of Euclidean dS$_3$ gravity. These probes are Wilson line operators, designed by exploiting the Chern-Simons formulation of 3D gravity. Our prescription uses non-unitary representations of $so(4)\simeq su(2)_L\times su(2)_R$, and we evaluate the Wilson line for states satisfying a singlet condition. We discuss how to reproduce the Green's functions of massive scalar fields in dS$_3$, the construction of bulk fields, and the quasinormal mode spectrum. We also discuss the interpretation of our construction in Lorentzian signature in the inflationary patch, via $SL(2,\mathbb{C})$ Chern-Simons theory.

 \vfill
\noindent \today

\end{titlepage}
}

\newpage

\tableofcontents

\newpage
\section{Introduction}

The Chern-Simons formulation of three-dimensional gravity seems more amenable to quantization than the more traditional metric formulation \cite{Achucarro:1987vz, Witten:1988hc}. One advantage is that the gauge theory formulation makes evident the topological nature of Einstein's theory in three dimensions.  Also, Chern-Simons theory has inherently  holographic properties: upon specifying a gauge group and boundary conditions on a 3-manifold with a boundary, the Chern-Simons theory can be viewed as dual to a conformal theory living on the boundary \cite{Moore:1989yh,Elitzur:1989nr,Verlinde:1989ua}. These features have propelled the use of the Chern-Simons formulation as a computational tool in perturbative gravity.

However, this alternative formulation of 3D gravity comes with a cost: local observables that are intuitive in a metric formulation---such as distances, surfaces, volumes, and local fields---are seemingly lost in Chern-Simons theory.  To reintroduce this intuition, Wilson lines present themselves as reasonable objects in Chern-Simons that could restore portions of our geometric and local intuition \cite{Castro:2018srf}. In the early stages, it was clear that a Wilson line anchored at the boundary would correspond to a conformal block in the boundary theory \cite{Witten:1988hf,Verlinde:1989ua}; more recently this proposal has been made more precise and explicit for $SL(2)$ Chern-Simons theory \cite{Alkalaev2015,Hijano2015a,Alkalaev2016,Besken2016,Fitzpatrick2017,Besken:2018zro,Hikida:2018dxe,DHoker:2019clx}. In the context of AdS$_3$ gravity, where the relevant gauge group is $SO(2,2)$, Wilson lines have been applied in a plethora of different contexts \cite{Witten:1989sx,Carlip:1989nz,Skagerstam:1989ti,deSousaGerbert:1990yp,Vaz:1994tm}, with recent applications ranging from the computation of holographic entanglement entropy \cite{Ammon:2013hba,deBoer:2013vca,Castro:2014mza,deBoer:2014sna,Hegde2016,Chen:2016uvu,Bhatta:2016hpz} to the probing of analytic properties of an eternal black hole \cite{Castro:2016ehj,Henneaux:2019sjx}.   Applications of Wilson lines in Chern-Simons to flat space holography includes \cite{Basu:2015evh}, and to ultra-relativistic cases \cite{Castro:2015csg,Azeyanagi:2018har}.

In the present work we will study $SO(4)$ Chern-Simons theory on a Euclidean compact manifold. This theory can be interpreted as a gravitational theory with positive cosmological constant, i.e. Euclidean dS$_3$ gravity.  This instance is interesting from a cosmological perspective, where Chern-Simons theory could provide insights into appropriate observables in quantum cosmology. It is also powerful, since there is an extensive list of exact results in Chern-Simons theory for compact gauge group. Previous efforts that exploited this direction of Chern-Simons theory as a toy model for quantum cosmology include \cite{Carlip:1992wg,Guadagnini:1995wv,Banados:1998tb, Park:1998yw, Govindarajan:2002ry,Castro:2011xb}.  

Our main emphasis is to interpret Wilson lines in $SO(4)$ Chern-Simons theory as local probes for dS$_3$ gravity, which follows closely the proposal in \cite{Castro:2018srf} for $SO(2,2)$ Chern-Simons. The basic idea is as follows. We will consider a connection $\mathscr{A}$ valued on $so(4)$, and a Wilson line stretching from a point $x_i$ to $x_f$:
 \be \label{eq:wilsonline1}
 W_{\mathscr{R}}(x_i,x_f) = \left<U_f \right| \mathscr{P}\mbox{exp}\left(-\int_{x_i}^{x_f}\mathscr{A}  \right)\left|U_i\right>~.
 \ee
 There are two important ingredients in defining this object. First we need to select a representation $\mathscr{R}$ of $so(4)$. This choice will encode the physical properties of the local probe, such as mass and spin. The second ingredient is to select the endpoint states $\left|U_{i,f}\right>$: the freedom in this choice encodes the gauge dependence of $W_{\mathscr{R}}(x_i,x_f)$. More importantly, their choice will allow us to relate $ W_{\mathscr{R}}(x_i,x_f)$ to the Euclidean Green's function of a massive field propagating on $S^3$. And while our choices are inspired by the analogous computations in AdS$_3$ gravity, they have a standing on their own. We will motivate and introduce the ingredients needed to have a interesting interpretation of \eqref{eq:wilsonline1} using solely $SO(4)$ Chern-Simons theory.

The interpretation of our results in the Euclidean theory will have its limitations if they are not analytically continued to Lorentzian signature. For example, recognising if the information contained in $W_{\mathscr{R}}(x_i,x_f)$ is compatible with causality necessitates a Lorentzian understanding of the theory. This is tied with the issue of bulk locality and reconstruction  in de Sitter, which remains intriguing in cosmological settings. In the Chern-Simons formulation, the Lorentzian theory corresponds to a theory with gauge group $SL(2,\mathbb{C})$. We will present the basics of how to discuss our results in $SL(2,\mathbb{C})$ Chern-Simons theory, and their relation to the Euclidean theory.  One interesting finding is that our choice of representation in $SO(4)$ Chern-Simons  naturally leads to quasinormal modes in the static patch of dS$_3$ when analytically continued.

\subsection{Overview}

In Sec.~\ref{sec:cs}, we review the Chern-Simons formulation of Euclidean dS$_3$ (EdS$_3$) gravity, establishing our conventions along the way.

In Sec.~\ref{sec:wl}, we describe Wilson lines in $SO(4)\simeq SU(2)\times SU(2)$ Chern-Simons. We show how the Green's function on EdS$_3$ of a scalar field  of given mass can be described by a Wilson line evaluated in a non-unitary representation of the algebra, which we construct in detail. These unusual representations of $su(2)$ resemble the usual spin-$l$ representation, with the important distinction that $-1<l<0$. And while it might be odd to treat $l$ as a continuous (negative) parameter, these features will be key to recover local properties we attribute to dS$_3$ in Chern-Simons theory.   

In Sec.~\ref{sec:local}, we take this further and show how this description of the Wilson line can be used to define local states in the geometry. We present a map between states in the algebraic formulation and the value of a corresponding scalar pseudofield in the metric formulation, and we build an explicit position-space representation of the basis states. We also match the action of the generators of the algebra to the Killing vectors of the geometry. The local pseudofields constructed from the Wilson line continue to quasinormal modes in the static patch, and they are acted on by an $sl(2,\mathbb{R})\times sl(2,\mathbb{R})$ inherited from our representations. This can be contrasted to a similar $sl(2,\mathbb{R})$ structure of the quasi-normal mode spectrum that was discovered and dubbed a ``hidden symmetry'' of the static patch in \cite{Anninos:2011af}.


In Sec.~\ref{sec:analyticcont}, we discuss how to analytically continue our results to Lorentzian dS$_3$ gravity, which is described by an $SL(2,\mathbb{C})$ Chern-Simons theory. We find that our exotic $so(4)$ representations analytically continue to a highest-weight representation of an $sl(2,\mathbb{R})\times sl(2, \mathbb{R})$ slice of $sl(2,\mathbb{C})$. 

In Sec.~\ref{sec:discussion}, we highlight our main findings and discuss future directions to further explore quantum aspects of dS$_3$ gravity. Finally, App.~\ref{app:conv} collects some of our conventions for easy reference, and App.~\ref{app:ds3} reviews some basic facts about the metric formulation of dS$_3$. In App.~\ref{app:cont}, we give more details about how to construct an analytic continuation between the $SO(4)$ and $SL(2,\mathbb{C})$ Chern-Simons theories.

\section{Chern-Simons formulation of Euclidean dS$_3$ gravity}\label{sec:cs}

For the purposes of setting up notation and conventions we begin with a short review of Chern-Simons gravity, focusing on its relation to Euclidean dS$_3$ gravity. This is based on the  original  formulation of $3$D gravity as a Chern-Simons theory \cite{Achucarro:1987vz, Witten:1988hc}; and related work on Euclidean dS$_3$ in the Chern-Simons formulation are \cite{Banados:1998tb,Castro:2011xb}, although we warn the reader that conventions there might be different than ours. In App.\,\ref{app:ds3} we provide a review of the metric formulation of dS$_3$ gravity. 

Consider Chern-Simons theory on $\mathcal M = S^3$ with gauge group $SO(4)$. This group manifestly splits into $ SO(4)\simeq SU(2)_L \times SU(2)_R$, and in terms of its Lie algebra we use generators $L_a$ for $su(2)_L$ and $\bar L_a$ for $su(2)_R$, $a=1,2,3$. Our conventions are such that
\be \label{eq:su2}
[L_a, L_b] = i \epsilon_{abc} L^c~, 
\ee
and similarly for the $\bar L_a$; we also set $\epsilon_{123}\equiv 1$.
There is an invariant bilinear form given by the trace: we take
\be \label{eq:bilinearsu2}
\mbox{Tr}(L_a L_b) =\mbox{Tr}(\bar L_a \bar L_b) = {1\over 2}\delta_{ab}~.
\ee
Indices in \eqref{eq:su2} are raised with $\delta^{ab}$.

The $SO(4)$ Chern-Simons action relevant for Euclidean dS$_3$ gravity is
\be 
S_{E} = S_{\rm CS}[A] - S_{\rm CS}[\bar A]~,\label{eq:CSaction}
\ee
where  
\be
A = A_\mu^a\,L_a \,\dd x^\mu~,\qquad  \bar A =  \bar A_\mu^a \,\bar L_a\, \dd x^\mu~, 
\ee
and the individual actions are
\be S_{\rm CS}[A] = -\frac{k}{4\pi} \int_{\mathcal M} \mbox{Tr}\left(A\wedge dA + \frac{2}{3} A \wedge A \wedge A\right) ~,\label{eq:SCS}
\ee
 and similarly for $S_{\rm CS}[\bar A]$.

The relation to the first-order formulation of the Einstein-Hilbert  action is as follows. The  algebra that describes the isometries of Euclidean dS$_3$ is
\begin{align}
[J_{ab}, P_c] &= -\delta_{ac} P_b + \delta_{bc} P_a~,\cr
[J_{ab},J_{cd}] &= -\delta_{ac} J_{bd} + \delta_{bc} J_{ad} + \delta_{ad} J_{bc} - \delta_{bd} J_{ac}~,\cr
[P_a,P_b] &=  -\Lambda J_{ab}~,\label{eq:so4algebra}
\end{align}
where $\Lambda = \frac{1}{\ell^2}$, and $\ell$ is the radius of the $3$-sphere. Here $P_a$ and $J_{ab}$ are the generators of translations and rotations of the ambient $\mathbb{R}^4$, respectively. We also raise indices with $\delta^{ab}$. It is convenient to define the dual
\be 
J_a = \frac{1}{2} \epsilon_{abc} J^{bc}~, \indent J_{ab} = \epsilon_{abc} J^c~.\label{eq:Jdefs}
\ee
In relation to the $su(2)$ generators, we identify
\be
L_a = -\frac{i}{2}(J_a + \ell\, P_a)~, \indent \bar L_a = -\frac{i}{2}(J_a - \ell \, P_a)~.\label{eq:LLbar}
\ee 
The variables in the gravitational theory are the vielbein and spin connection,
\be 
e^a = e_\mu^{\ a} \dd x^\mu~, \indent \omega^a = \frac{1}{2} \epsilon^a_{\ bc} \omega_\mu^{\ bc} \dd x^\mu~. 
\ee
The vielbein is related to the metric as $g_{\mu\nu} = e_\mu^{\ a} e_\nu^{\ b} \delta_{ab}$.
We define the gauge field in terms of these geometrical variables as
\be 
A = i\left(\omega^a + \frac{1}{\ell}e^a\right) L_a~, \indent \bar A= i\left(\omega^a - \frac{1}{\ell}e^a\right) \bar L_a~.\label{eq:aabar}
\ee
Using \eqref{eq:aabar}, the action \eqref{eq:CSaction} becomes
\be \label{eq:euceh}
S_{E} = \frac{k}{2\pi\ell} \int_{\mathcal M} \left[e^a \wedge (d\omega_a - \frac{1}{2}\epsilon_{abc} \omega^b \wedge \omega^c) - \frac{1}{6\ell^2} \epsilon_{abc} e^a\wedge e^b\wedge e^c\right]~,
\ee
which reduces the Einstein-Hilbert action with positive cosmological constant given the identification
\be 
k = \frac{\ell}{4G_3}~.
\ee
The equations of motion from \eqref{eq:CSaction} simply give the flatness condition,
\be 
F = dA + A \wedge A = 0~,\qquad \bar F = d\bar A + \bar A \wedge \bar A = 0~,
\ee
which are related to the Cartan and Einstein equation derived from \eqref{eq:euceh} after using \eqref{eq:aabar}.

The background we will mostly focus on is $S^3$, which we will cast as
\begin{equation}
{\dd s^2\over \ell^2}=\dd r^2+\cos^2 r \dd \tau^2 + \sin^2 r \dd\phi^2~,
\end{equation}
with $(\tau, \phi)\sim (\tau,\phi)+ 2\pi(m,n)$ and $m,n\in\ZZ$; see App.\,\ref{app:coord} for further properties of this background.  In the Chern-Simons language, the associated connections that reproduce the vielbein and spin connection are
\begin{align}
A&=i L_2 \dd r + i\le(L_3 \cos r + L_1 \sin r\ri)\le(\dd \phi+\dd \tau\ri) ~,\cr
\bar A&=-i L_2 \dd r + i\le( L_3 \cos r -  L_1 \sin r\ri)\le(\dd \phi-\dd \tau\ri)~.
\end{align}
Note that we are using the same basis of $su(2)$ generators for both $A$ and $\bar A$. This is convenient since we then can read off the metric as
\be
g_{\mu \nu} =- {\ell^2\over 2} {\rm Tr}\le[ \le( A_\mu -\bar A_\mu\ri)\le( A_\nu - \bar A_\nu\ri)\ri]~.
\ee
The corresponding group elements that we will associate to each flat connection read\footnote{The notation $\tilde g_R$ here will be justified and explained in Sec.\,\ref{sec:wilsongreen}.}
\begin{align}\label{eq:s3bckg1}
A&= g_L \dd g_L^{-1}~,\qquad g_L=e^{-irL_2}e^{-i (\phi+\tau)L_3}~, \cr
\bar A &= \tilde g_R^{-1}\dd \tilde g_R~,\qquad \tilde g_R=e^{i(\phi-\tau) L_3}e^{-ir L_2}~.
\end{align}
 This can be checked explicitly by using the following corollary of the Baker-Campbell-Hausdorff formula,
\begin{equation}
e^{-i \alpha L_a} L_b e^{i\alpha L_a}=\cos(\alpha)L_b+ \sin(  \alpha)\epsilon_{abc}L^c~. \label{eq:bch}
\end{equation}

\section{Wilson lines in $SO(4)$ Chern-Simons }\label{sec:wl}
A gauge-invariant observable in Chern-Simons theory is the Wilson loop operator, which in the Euclidean theory with gauge group $SO(4) \simeq SU(2)_L\times SU(2)_R$  reads
\be 
W_{\mathscr{R}}(C) = \mbox{Tr}_{\mathscr{R}} \left(\mathscr{P}\mbox{exp}\left(-\oint_{C} A\right)\mathscr{P}\mbox{exp}\left(-\oint_{C} \bar A\right)\right)~,
\ee
where $C$ is a closed loop in the 3-manifold ${\cal M}$. Here $\mathscr R$ is a particular representation of the Lie algebra associated to the Chern-Simons gauge group. One of the challenges of the Chern-Simons formulation of 3D gravity is to build local probes in a theory that insists on being topological. Here we will design those probes by considering a Wilson \emph{line} operator, i.e., we will be interested in 
 \be \label{eq:wilsonline}
 W_{\mathscr{R}}(x_i, x_f) = \left<U_f \right| \mathscr{P}\mbox{exp}\left(-\int_\gamma A\right)\mathscr{P}\mbox{exp}\left(-\int_\gamma \bar A\right)\left|U_i\right>~.
 \ee
 The curve $\gamma(s)$ is no longer closed but has endpoints $x_i, x_f$. This operator is no longer gauge-invariant, which is reflected in the fact that we need to specify  states at its endpoint, denoted as $\left|U_i\right>, \left|U_f\right>$. In the following we will discuss representations $\mathscr R$ of $so(4)$, and suitable endpoint states, giving $ W_{\mathscr{R}}(x_i, x_f) $ local properties we can naturally relate to the metric formulation. The representations we consider will differ from the unitary representations that are typically considered in $SU(2)$ Chern-Simons theory.\footnote{In the semi-classical regime, there is no principle in Chern-Simons theory that favors a choice of one representation over another. The choices we make, both for the representations and endpoint states, allow us to reproduce gravitational observables in de Sitter spacetime.}

Our strategy to select the representation and the endpoint states is inspired by the proposal in \cite{Ammon:2013hba,Castro:2018srf}, which is a prescription to use Wilson lines as local probes in AdS$_3$ gravity.  The basic observation is to view $W_{\mathscr{R}}(x_i,x_f)$ as the path integral of a charged (massive) point particle. In this context the representation  ${\mathscr{R}}$ parametrizes the Hilbert space for the particle, with the Casimir of ${\mathscr{R}}$ carrying information about the mass and spin (i.e., quantum numbers) of the particle \cite{Witten:1989sx,Carlip:1989nz,Beasley:2009mb}. With this perspective, our first input is to consider representations of $so(4)$ that carry a continuous parameter that we can identify with the mass of particle. As we will show in the following, this requirement will force us to consider non-unitary representations of the group which we will carefully construct. 

In the subsequent computations we will leave the connections $A$ and $\bar A$ fixed, and quantize appropriately the point particle for our choice of ${\mathscr{R}}$. From this perspective, $W_{\mathscr{R}}(x_i,x_f)$ captures how the probe is affected by a given background characterized by $A$ and $\bar A$. Here is where our choice of endpoint states will be crucial: our aim is to select states in ${\mathscr{R}}$ that are invariant under a subgroup of $so(4)$. Selecting this subgroup appropriately will lead to a novel way of casting local fields in the Chern-Simons formulation of dS$_3$ gravity.

\subsection{Non-unitary representations of $so(4)$}\label{sec:reps}

Since $so(4)\simeq su(2)_L\times su(2)_R$, let us focus first on a single copy of $su(2)$.  Recall that in our conventions, the $su(2)$ generators satisfy the algebra \eqref{eq:su2}.
The unique Casimir operator is the quadratic combination\footnote{Note that this definition of the Casimir discards the overall normalization of the bilinear form in \eqref{eq:bilinearsu2}.}
\be 
L^2 = L_1^2 + L_2^2 + L_3^2~.
\ee
We can build raising and lowering operators by defining
\begin{align} \label{eq:newbasis}
L_{\pm } \equiv L_1 \pm i L_2~,\qquad L_0 \equiv L_3~.
\end{align}
For a compact group like $su(2)$ all unitary representations are finite dimensional and labelled by a fixed (half-)integer, the spin, as in the usual $SU(2)$ Chern-Simons theory.  To introduce a continuous parameter, we need to build representations that are more analogous to the infinite-dimensional, highest-weight representations of $sl(2,\mathbb{R})$.  This forces us to consider non-unitary representations, nevertheless a natural choice to make contact with local fields in dS$_3$, as will show.  

For unitary representations we would have that all of the $L_a$'s are Hermitian. Here we will relax this condition and choose generators that are not necessarily Hermitian. In particular,  a consistent choice for a non-unitary representation that respects the Lie algebra is to take $L_1, L_2$ to be anti-Hermitian and $L_3$ to be Hermitian, which results in
\be 
L_{\pm }^\dagger = -L_{\mp }~,\indent L_0^\dagger = L_0~.\label{eq:Hermiticity}
\ee
While it is not unique, this is the choice we will use to build a non-unitary representation. Notice that it is inconsistent to take all the generators to be anti-Hermitian, as this would violate the commutation relations \eqref{eq:su2}. Informally, we only modify the construction of su(2) representations as much as needed to obtain a continuous Casimir. As we see below, this modification is sufficient to obtain that property, with the rest of the construction mirroring the usual unitary case.

Our representation, despite its lack of unitarity, has to satisfy some minimal requirements which we will now discuss. We have a basis of vectors (states) that are joint eigenstates of $L^2$ and $L_0$. These are denoted $|l,p\rangle$ with
\begin{align}
L^2|l,p\rangle&= c_2(l)|l,p\rangle~, \label{eq:L2eigen} \\
L_0|l,p\rangle&=(l-p)|l,p\rangle~.
\end{align}
Here $l$ labels the representation, i.e. controls the quadratic Casimir $c_2(l)$, and $p$ labels the $L_0$ eigenvalue. Note that in a unitary representation we would use $m=l-p$, but we will find it more useful to use $p$ as a label. We seek to build a representation such that the spectrum of $L_0$ is bounded (either from above or below), and that the norm squared of the states $|l,p\rangle$ is positive. To achieve these requirements, we build a representation by introducing a {\it highest weight state}. We define this state as
\be 
L_0 \left| l,0\right> = l\left| l,0\right>~, \indent L_{+} \left| l,0\right> = 0~.
\ee
This in particular implies that we will create states by acting with $L_-$ on $\left| l,0\right>$, and hence a basis for eigenstates is schematically given by $|l,p\rangle \sim (L_-)^p\left| l,0\right>$ with $p$ a positive integer.\footnote{This follows from the commutation relation between $L_\pm$ and $L_0$.}  

Next we need to ensure that the norm of these states is {\it positive}; this will impose restrictions on the Casimir, and hence $l$.  A useful identity in this regard is
\begin{align} 
\nn |L_{\pm } \left|l,p\right>|^2 &= - \left<l,p\right| L_{\mp} L_{\pm} \left| l,p\right>\\
&= - c_2(l)+(l-p)(l-p\pm1)~.\label{eq:Lpnorm}
\end{align}
The minus sign in the first line comes from anti-Hermiticity in \eqref{eq:Hermiticity}. In going from the first to the second line we used $L_{\mp } L_{\pm } = L^2 - L_0^2 \mp L_0$.  The norm of $L_{+ } \left|l,0\right>$ vanishing gives 
\be\label{eq:Casvalue}
c_2(l)=l(l+1)~,
\ee
relating the label $l$ with the Casimir of the representation. Positivity of  the norm of the first descendant requires
 \be
 \nn |L_{-} \left|l,0\right>|^2 = -2l >0~,
\ee
which clearly dictates that $l$ is strictly negative. Any other state in the representation will be of the form
\be\label{eq:eigen1}
\left|l,p\right> = \mathfrak{c}_p (L_-)^p\left| l,0\right>~,
\ee
where the normalization $\mathfrak{c}_p$ is adjusted such that 
\be\label{eq:eigen2}
\left<l,p' | l,p\right> = \delta_{p',p}~, \qquad p=0,1,2,\cdots~.
\ee
Demanding this relation leads to 
\begin{align}
L_{+} \left|l, p\right> &= -\sqrt{p(p-2l-1)}\left|l, p-1\right>~,\label{eq:Lplusoperation}\\
L_{-} \left|l, p\right> &= \sqrt{(p+1)(p-2l)}\left|l, p+1\right>~.\label{eq:Lminusoperation}
\end{align}
 The fact that the roles of the raising and lowering operators appear flipped, in other words $L_+$ lowers and $L_-$ raises $p$, simply results from our convention in \eqref{eq:L2eigen}. If we had labelled states by their eigenvalue of $L_0$ they would raise and lower in the same way as the usual unitary $sl(2,\mathbb{R})$ representations. The minus sign in \eqref{eq:Lplusoperation} is more fundamental. It was not present in highest weight representations of $sl(2,\mathbb{R})$; here it  is necessary for the action of $L_{\pm}$ to be consistent with the $su(2)$ commutation relations.\footnote{Normalization only determines $L_{\pm}|l,p\rangle$ up to a phase.} 

In the unitary case, representations are finite-dimensional since there is an upper bound for $p$.  Additionally, the Casimir is strictly positive, and $l$ is constrained to be either integer or half-integer. These constraints all come from demanding the positivity of squared norms. For our non-unitary representations, relaxing the requirement of Hermiticity means that $p$ is not bounded and the Casimir is not necessarily positive.
Our choices also lead to a spectrum of $L_0$ unbounded from below, whose eigenstates are \eqref{eq:eigen1}-\eqref{eq:eigen2}. We also note that the Casimir is allowed to be negative since $l<0$; in particular, for the range $-1< l < 0$ we have
\be -\frac{1}{2} < c_2 < 0~. \ee

Our representation has a well-defined character too. Suppose we have a group element $M \in SU(2)$ which can be decomposed as
\begin{equation}
M=V^{-1} e^{i\alpha L_0} V~.
\end{equation}
Its character is simply given by
\begin{align}\label{eq:ch123}
\text{Tr}(M)=\sum_{p=0}^{\infty} \langle l,p|e^{i\alpha L_0}|l,p\rangle =\frac{e^{i\alpha(l+1)}}{e^{i\alpha}-1}~.
\end{align}

Finally, notice that for a fixed Casimir there are actually two distinct representations labelled by the two solutions for $l$ in \eqref{eq:Casvalue}. These solutions are
\be\label{eq:choice}
l_\pm = -\frac{1\pm \sqrt{1+4 c_2}}{2}~.
\ee
One representation has $-1<l_+<-{1\over2}$ while the other has $-{1\over 2}<l_-<0$, and each of these representations  will be coined as ${\mathscr{R}}_\pm$. The role of ${\mathscr{R}}_\pm$ will become important later, when we compare the Wilson line to the Euclidean Green's function, and in the construction of local pseudofields. In particular, we will see that both representations are necessary to generate a complete basis of solutions for local fields in dS$_3$.

\subsubsection{Singlet states}\label{sec:singlet}
Returning to $so(4)\simeq su(2)_L\times su(2)_R$, let's add a set of operators $\bar L_a$ with the same commutation relations as the unbarred ones and which commute with them: 
\be [L_a, \bar L_b] = 0~.\ee
In the following we will be interested in building a state $|U\rangle$, assembled from the non-unitary representations of $su(2)$, that is invariant under a subset of the generators in $so(4)$. These states, denoted {\it singlet states}, will serve as endpoint states which we will use to evaluate the Wilson line \eqref{eq:wilsonline}. This construction is motivated by the derivations for $so(2,2)\simeq sl(2,\mathbb{R})_L\times sl(2,\mathbb{R})_R$ presented in \cite{Castro:2018srf}. Here we will review the derivation as presented there, adapted appropriately to $so(4)$. 

Singlet states of $so(4)$ can be constructed as follows.  Consider a group element $U \in SU(2)$, and define the rotated linear combination
\be Q_a(U) = L_a  + D_a^{\ a'}(U) \bar L_{a'}~,\ee
where $D_a^{\ a'}$ corresponds to the adjoint action of the group; see App.\,\ref{app:conv} for our conventions.
We define a state $\left|U\right>$ through its annihilation by $Q_a(U)$,
\be Q_a(U) \left|U\right> = 0~. \label{eq:Qdef}\ee
In other words, $\left|U\right>$ is a state that is invariant under a linear combination of $so(4)$ generators specified by $Q_a(U)$. This equation is crucial: the inclusion of both copies of $su(2)$ will ensure that the states $\left|U\right>$ will prevent a factorization in our observables, and will allow us to interpret our choices in the metric formulation.

There are two interesting choices of $|U\rangle$  for which it is useful to build explicit solutions to \eqref{eq:Qdef}.  We refer to our first choice as an  {\it Ishibashi} state: it is defined by selecting a group element $U=\Sigma_{\rm Ish}$ such that 
 \be
 D_k^{\ k'}(\Sigma_{\rm Ish})\, L_{k'} = \Sigma_{\rm Ish}\, L_k \, \Sigma_{\rm Ish}^{-1} = -L_{-k}~,\label{eq:Ishdef}
 \ee
 where we are using the basis \eqref{eq:newbasis}, and therefore $k=-,0,+$. 
The corresponding group element is 
\be
\Sigma_{\rm Ish} =  e^{\frac{\pi}{2}(L_{+} - L_{-})}=e^{i \pi L_2} ~.\label{Ishgroup}
\ee
The corresponding singlet state, i.e., Ishibashi state, is the solution to
\be\label{eq:ishicond}
(L_k - \bar L_{-k}) \left|\Sigma_{\rm Ish}\right> = 0~.
\ee
This equation has a non-trivial solution for  the non-unitary representations built in Sec.\,\ref{sec:reps}. Consider the basis of states in \eqref{eq:eigen1}-\eqref{eq:eigen2} for each copy of $su(2)$ of the form
\be
\sum_{p,\bar p} a_{p,\bar p}\left|l, p\right> \otimes \left|\bar l, \bar p\right> ~,
\ee
with coefficients $a_{p,\bar p}$, as an ansatz for $\left|\Sigma_{\rm Ish}\right> $. The $k=0$ condition in \eqref{eq:ishicond} sets $l=\bar l$, and $k=\pm$ will give $a_{p,\bar p}= (-1)^p\delta_{p,\bar p}$, up to an overall normalization independent of $p$. The resulting state is
\be\label{eq:sish}
\left|\Sigma_{\rm Ish}\right> = \sum_{p=0}^\infty \left (-1)^p |l,p,p\right>~,
\ee
where $ \left|l, p, \bar p\right>\equiv \left|l, p\right> \otimes \left|l, \bar p\right>.$
 
The second choice will be coined {\it crosscap} state.  In this instance, we select $U=\Sigma_{\rm cross}$ such that 
\be
D_k^{\ k'}(\Sigma_{\rm cross})\, L_{k'} = \Sigma_{\rm cross}\, L_k\, \Sigma_{\rm cross}^{-1} = -(-1)^k L_{-k}~,\label{eq:crossdef}
\ee
which leads to the group element 
\begin{align}
\Sigma_{\rm cross} = e^{\frac{i \pi}{2}(L_{+} + L_{-})}=e^{i \pi L_1} ~.\label{crossgroup}
\end{align}
Using \eqref{eq:crossdef} in \eqref{eq:Qdef}, the crosscap state satisfies 
\begin{align}
(L_ k- (-1)^k \bar L_{-k}) \left|\Sigma_{\rm cross}\right> = 0~,\label{eq:crosscond}
\end{align}
and in terms of the non-unitary $su(2)$ representations  the solution to these conditions are
\begin{align} \label{eq:scross}
\left|\Sigma_{\rm cross}\right> = \sum_{p=0}^\infty \left|l,p,p\right>~.
\end{align}

In contrast to the Virasoro construction, it is important to emphasise that here we don't have an interpretation of \eqref{eq:ishicond} and \eqref{eq:crosscond} as a boundary condition of an operator in a CFT$_2$ as in \cite{Ishibashi:1988kg,Cardy:2004hm}. We are using (and abusing) the nomenclature used there because of the resemblance of \eqref{eq:ishicond} and \eqref{eq:crosscond} with the CFT$_2$ conditions, and its close relation to the $so(2,2)$ states used in \cite{Castro:2018srf}. In this regard, it is useful to highlight some similarities and key differences in $so(4)$ relative to $so(2,2)$. A similarity is that our choice to use $p$ rather than the eigenvalue of $L_0$ to label the states in the non-unitary representation was precisely motivated to make the states match with those in $sl(2,\mathbb{R})$. However, one difference is that   the group elements  \eqref{Ishgroup} and \eqref{crossgroup} differ by a factor of $i$ in the exponent compared to their $sl(2,\mathbb{R})$ counterparts in \cite{Castro:2018srf}. Also we  note that, unlike in the $sl(2,\mathbb{R})$ case, the relative phase in the state now appears in the Ishibashi state rather than the crosscap state. This is due to the extra minus sign in \eqref{eq:Lplusoperation}.

Another important property of the singlet states is their transformation under the action of $SU(2)$ group elements. Consider $G(L)\in SU(2)_L$, and $\bar G(R^{-1})\in SU(2)_R$ for each copy appearing in $SO(4)$. A simple manipulation shows that
\be
G(L) \bar G(R^{-1}) Q_a(U)\left|U\right> = D_a^{~a'}(L^{-1}) Q_a'(LUR) G(L) \bar G(R^{-1})\left|U\right>=0~.
\ee
Thus we have
\be \label{eq:rotationsig}
G(L) \bar G(R^{-1}) \left|U\right> = \left|LUR\right>~.
\ee
This identity will be used heavily in the following derivations.  

\subsection{Wilson line and the Green's function}\label{sec:wilsongreen}

We now come back to evaluating the Wilson line \eqref{eq:wilsonline}. We select as endpoints states
\be
\left|U_i\right>=\left|U_f\right>=\left|\Sigma\right>~,
\ee
with the choice of $\left|\Sigma\right>$ being either
\be
\left|\Sigma_{\rm Ish}\right> ~ {\rm or} \left|\Sigma_{\rm cross}\right>~,
\ee
i.e., the Ishibashi \eqref{eq:sish} or crosscap \eqref{eq:scross} state. From this perspective we can view \eqref{eq:wilsonline} as
\be\label{eq:innerp1}
W_{\mathscr{R}}(x_i, x_f) = \langle \Sigma |G(L) \bar G(R^{-1}) \left|\Sigma\right>~,
\ee
where we identify
\be
G(L) = \mathscr{P}\mbox{exp}\left(-\int_\gamma A\right)~,\qquad \bar G(R^{-1}) =\mathscr{P}\mbox{exp}\left(-\int_\gamma \bar A\right)~. 
\ee 
Given the properties of our singlet states, we can easily evaluate \eqref{eq:innerp1} as follows,
\begin{align}\label{eq:derivation}
W_{\mathscr{R}}(x_i, x_f) &= \langle \Sigma |G(L) \bar G(R^{-1}) \left|\Sigma\right>\cr
&= \langle \Sigma |G(L \, \tilde R) \left|\Sigma\right>\cr
&= \sum_{p=0}^\infty \langle l,p|G( L \, \tilde R)|l,p\rangle\cr
&= \frac{e^{i\alpha(l+1)}}{e^{i\alpha}-1}~. 
\end{align}
In the second line we used \eqref{eq:rotationsig} to move the right group element $R$ to the left, where 
\be
\tilde R \equiv \Sigma \, R \, \Sigma^{-1}~.
\ee
To obtain the third line in \eqref{eq:derivation} we use the explicit form of the states given by \eqref{eq:sish} and \eqref{eq:scross}, where both the Ishibashi and cross cap state report the same answer. Finally in the last equality we used the formula for the character in \eqref{eq:ch123}, where $\alpha$ in this case is defined via the equation
\begin{equation}\label{eq:alpha1}
L \Sigma \, R \, \Sigma^{-1}=V^{-1} e^{i\alpha L_0} V~,
\end{equation}
i.e., assuming we can diagonalise the left hand side, $\alpha$ captures  the eigenvalue of the group element in the inner product.

The interpretation of \eqref{eq:derivation} in the metric formulation of dS$_3$ gravity is interesting. First, we observe that for a pair of $su(2)$ Chern-Simons connections, 
\begin{align}\label{eq:flat2}
A= g_L \dd g_L^{-1}~, \qquad\bar A = g_R^{-1}\dd g_R~,
\end{align}
we have 
\be\label{eq:id234}
G(L) =g_L(x_f) g_L(x_i)^{-1}~, \qquad \bar G(R^{-1})= g_R(x_f)^{-1} g_R(x_i)~,
\ee
where we evaluated the path ordered integral for a path $\gamma$ with endpoints $(x_i,x_f)$. For concreteness, we will make the choice
\be\label{eq:sphere231}
g_L=e^{-irL_2}e^{-i (\phi+\tau)L_3}~,\qquad  \tilde g_R\equiv\Sigma \, g_R \, \Sigma^{-1} =e^{i(\phi-\tau) L_3}e^{-ir  L_2}~,
\ee
which for $SU(2)_L$ is the group element associated to $S^3$ in \eqref{eq:s3bckg1}. But it is important to stress that, with some insight, we are specifying $ \tilde g_R$ rather than $g_R$, since this is all we need at this stage to evaluate $W_{\mathscr{R}}(x_i, x_f)$. Using  \eqref{eq:sphere231}  we find that the solution for $\alpha$ in \eqref{eq:alpha1} is
\be
\cos\left(\frac{\alpha}{2}\right)=\cos(r_f)\cos(r_i)\cos(\tau_f-\tau_i)+\sin(r_f)\sin(r_i)\cos(\phi_f-\phi_i)~.
\ee
$\alpha$, which labels the equivalence class of $L \Sigma R \Sigma^{-1}$, can then be related to the geodesic distance between points $(x_i,x_f)$ on $S^3$ (see \eqref{eq:distace}):
\be
\alpha=\pm 2\Theta + 4\pi n~, \qquad n\in \ZZ ~,
\ee
with $n$ accounting for winding.
As explained in App. ~\ref{sect:prop}, the propagator of a scalar field of mass $m$ in dS$_3$ can be written as
\begin{align}
G(\Theta)&=\mathcal{G}_h(\Theta)+\mathcal{G}_{1-h}(\Theta)~,  \cr
\mathcal{G}_h(\Theta)&= a_h\frac{e^{-2ih\Theta}}{e^{-2i\Theta}-1}~, \label{eq:curly-g}
\end{align}
with 
\begin{equation}
a_h=\frac{i}{2\pi \ell} \frac{1}{1-e^{-4\pi i h}}~,\qquad h=\frac{1+\sqrt{1-(m\ell)^2}}{2}~.
\end{equation}
Equations (\ref{eq:derivation}) and (\ref{eq:curly-g}) lead us to conclude that if we pick a representation $\mathscr{R}=\mathscr{R}_+$ in \eqref{eq:choice} with $l=-h$ then 
\begin{equation}
W_{\mathscr{R}_+}(x_i, x_f)=\frac{1}{a_h} \mathcal{G}_h(\Theta)~.
\end{equation}
Similarly, picking instead a representation $\mathscr{R}=\mathscr{R}_-$ in \eqref{eq:choice}, where now $l=h-1$, leads to
\begin{equation}
W_{\mathscr{R}_-}(x_i, x_f)=\frac{1}{a_{1-h}} \mathcal{G}_{1-h}(\Theta)~.
\end{equation}
The full propagator can then be written as
\begin{equation}\label{eq:eucgreen111}
G(\Theta)=a_h W_{\mathscr{R}_+}(x_i, x_f) + a_{1-h} W_{\mathscr{R}_-}(x_i, x_f)~.
\end{equation}
$\mathscr{R}_\pm$ are the two possible representations with the same Casimir $c_2=h(h-1)=-\frac{m^2 \ell^2}{4}$. We emphasize that, unlike in AdS$_3$, we need to consider both of these representations to obtain the correct propagator. This is related to the fact that the de Sitter propagator is not simply given by the analytic continuation from AdS$_3$ due to differences in causal structures~\cite{Bousso:2001mw,Harlow:2011ke}.

Moving away from the specificity of group elements \eqref{eq:sphere231}, for any pair of flat connections \eqref{eq:flat2}, we will have that $W_{\mathscr{R}_\pm}(x_i, x_f)$ gives the $\mathcal{G}_{h,1-h}(\Theta)$ contribution to the Green's function between the points $(x_i, x_f)$ in  the Euclidean space with metric 
\be
g_{\mu \nu} =- {\ell^2\over 2} {\rm Tr}\le[ \le( A_\mu -\Sigma\, \bar A_\mu\, \Sigma^{-1}\ri)\le( A_\nu -\Sigma\, \bar A_\nu\, \Sigma^{-1}\ri)\ri]~.
\ee
A proof of this statement, beyond the explicit computation done here for $S^3$, follows step by step the derivations in \cite{Castro:2018srf} for $so(2,2)$ adapted to $so(4)$. The geometric role of our singlet states is now more clear: $\Sigma$ is the group element that controls how the right connection $\bar A$ acts as a left element relative to $A$, and vice-versa. These derivations also establish the gravitational Wilson line as a local probe of the Euclidean dS$_3$ geometry, and hence will allow us to investigate notions of locality in the Chern-Simons formulation of gravity.

\section{Local pseudofields from Wilson lines}\label{sec:local}

The aim of this section is to further extract local quantities from the gravitational Wilson line. We will focus on the background connections associated to the round 3-sphere for concreteness, and show how to build local pseudofields from the singlet states used in the previous section. We use the term  ``pseudofields'' because while the objects we will build from a single irreducible representation $\mathscr{R}$ (either $\mathscr{R}_+$ or $\mathscr{R}_-$) are local, and behave in many ways like fields, both representations are needed to form a complete basis for local fields in dS$_3$.

\subsection{Wilson line as an overlap of states}\label{sec:overlap}

Until now, we have described the Wilson line $W_{\mathscr{R}}(x_i, x_f)$  as the diagonal matrix element of an operator in a singlet state, as done in \eqref{eq:innerp1}. For the purpose of building local probes, we want to rewrite this  operator as a suitable overlap between states. From \eqref{eq:id234} we can write \eqref{eq:innerp1} as
\be
W_{\mathscr{R}}(x_i, x_f) = \langle \Sigma|\bar{G}(g_R(x_f)^{-1})G(g_L(x_f))\,  G(g_L(x_i)^{-1})  \bar{G}(g_R(x_i)) |\Sigma \rangle~.
\ee
If our representation $\mathscr{R}$ used Hermitian generators, we would simply note that for unitary group elements, i.e.,  
\be
g_R^{-1} = g_R^\dagger~, \qquad g_L^{-1} = g_L^\dagger~,
\ee
 we would have $W_{\mathscr{R}}(x_i, x_f)=  \langle U (x_f)| U(x_i)\rangle$ with $| U(x)\rangle=G(g_L(x)^{-1}) \bar{G}(g_R(x)) |\Sigma \rangle$. However, our representation is {\it non}-unitary, and hence these manipulations require some care. 
 
 Define the following state:
\begin{equation}\label{eq:ufield}
|U(x)\rangle=G(g_L(x)^{-1}) \bar{G}(g_R(x))|\Sigma\rangle~,
\end{equation}
We will focus exclusively on the background introduced in \eqref{eq:s3bckg1}. Because the representation we are using is non-unitary, we have
\begin{equation}
g_{L}(\tau,r,\phi)^\dagger=g_{L}(\tau,-r,\phi)^{-1} = g_{L}(\tau,r,\phi +\pi)^{-1} e^{i\pi L_3}~,
\end{equation}
and the same relation for $g_R$, which allow us to write  the Wilson line  as
\begin{equation}\label{eq:overlap1}
W_{\mathscr{R}}(x_i, x_f) = \langle U(\tau_f,r_f,\phi_f+\pi)|U(\tau_i,r_i,\phi_i)\rangle ~.
\end{equation}
In this equality we used
\be
e^{i\pi L_3} e^{i\pi \bar L_3} |\Sigma\rangle \sim  |\Sigma\rangle~,
\ee
since both singlet states are annihilated by $Q_a(\Sigma)$.

\subsection{Construction of local basis}\label{sec:localbasis}
Having written $W_{\mathscr{R}}(x_i, x_f)$  as an overlap of states, we now can start the process of defining a local pseudofield from $|U(x)\rangle$. The most natural way to split \eqref{eq:overlap1} is as done in \eqref{eq:ufield}. Still this has its inherent ambiguities: in defining $|U(x)\rangle$ we are splitting the cutting curve $\gamma(s)$ at some midpoint $x_0$, the choice of which is a gauge freedom at our disposal. More concretely, a general definition of the state should be  
\begin{equation}\label{eq:localchoice}
|U(x)\rangle=G(g_L(x_0)g_L(x)^{-1})\bar{G}(g_R(x_0)^{-1}g_R(x))|\Sigma\rangle
\end{equation}
where we restored the dependence on this midpoint split.  At this stage it is not clear to us that one choice of $g_{L,R}(x_0)$ is better than any other, so for sake of simplicity we will select $g_{L,R}(x_0)=\mathds{1}$, i.e. the identity element. Therefore we will be working with  \eqref{eq:ufield}, and explore the local properties of $|U(x)\rangle$.

First, we expand $|U(x)\rangle$ in the eigenstate $|l,p,\bar{p}\rangle$ basis:
\be |U(x)\rangle = \sum_{p,\bar p=0}^\infty \Phi^*_{p, \bar p}(x) |l, p, \bar p\rangle~,\ee
which we can reverse as
\begin{align}
\Phi_{p, \bar p}(x) &= \langle U(x)| l, p, \bar p\rangle~.
\end{align} 
$\Phi_{p, \bar p}(x)$ will be our basis of local pseudofields that will support the local properties in $ |U(x)\rangle$. To build this basis of eigenfunctions, we can translate the action of the generators $L_a$ on the basis vectors into the action of differential operators $\zeta_a$ acting on $\Phi_{p,\bar p}$. Specifically, we will find $\zeta_a, \bar{\zeta_a}$ such that
\begin{align} 
\langle U(x) | L_a | l, p, \bar p\rangle =\zeta_a \langle U(x)| l, p, \bar p\rangle~, \label{eq:diffopL-e}\\
\langle U(x) | \bar L_a | l, p, \bar p\rangle =  \bar{\zeta_a} \langle U(x)| l, p, \bar p\rangle~. \label{eq:diffopLbar-e}
\end{align}
Using \eqref{eq:Lplusoperation} and \eqref{eq:Lminusoperation}, the differential operators must therefore satisfy 
\begin{align}
\zeta_{+} \Phi_{p,\bar{p}}&=-\sqrt{p(p-2l-1)} \Phi_{p-1,\bar{p}} ~ \label{eq:Lplusop}\\
\zeta_{-} \Phi_{p,\bar{p}}&=\sqrt{(p+1)(p-2l)} \Phi_{p+1,\bar{p}} ~\label{eq:Lminusop}\\
\zeta_0 \Phi_{p,\bar{p}}&=(l-p)\Phi_{p,\bar{p}} ~,\label{eq:L0op}
\end{align}
and similarly for the barred sector.
It follows that $\Phi_{p, \bar p}$ satisfies the Casimir equation,
\begin{equation}\label{eq:KGphiE}
\left(\nabla^2+\bar\nabla^2\right) \Phi_{p,\bar{p}}(x)=2l(l+1)\Phi_{p,\bar{p}}(x)~.
\end{equation}
where $\nabla^2 = \delta^{ab}\zeta_a\zeta_b$, and $\bar\nabla^2 = \delta^{ab}\bar \zeta_a\bar\zeta_b$.\footnote{We will find that $\nabla^2+\bar{\nabla}^2=-\frac{1}{2}\nabla^2_{S^3}$, where $\nabla^2_{S^3}$ is the ordinary Laplacian for EdS$_3$.} Our strategy will be to build the differential operators for $(\zeta_a,\bar\zeta_a)$ based on \eqref{eq:diffopL-e}-\eqref{eq:diffopLbar-e}, and then solve for $\Phi_{p,\bar{p}}(x)$ from the differential equations \eqref{eq:Lplusop}-\eqref{eq:KGphiE}.

We will start by building the generators $\zeta_a$ for Euclidean dS$_3$. It is convenient to cast the state in \eqref{eq:ufield} as
\begin{align}
|U(x)\rangle&=G(g_L(x)^{-1}) \bar{G}(g_R(x))|\Sigma\rangle\cr
&=G(g_L(x)^{-1} \tilde g_R(x)^{-1})|\Sigma\rangle\cr
&=e^{i (\phi + \tau)L_3} e^{2irL_2}e^{-i (\phi-\tau) L_3}|\Sigma\rangle~. \label{eq:u-unbarred} 
\end{align}
In the second line we moved all the group elements to left, as in \eqref{eq:derivation}, and in the third line we used \eqref{eq:sphere231}.  Next, consider the action of partial derivatives on $\Phi_{p\bar{p}}(x)=\langle U(x)|l,p,\bar{p}\rangle$:
\begin{align}
\partial_+ \langle U(x)|l,p,\bar{p}\rangle &= -i \langle U(x)|L_3|l,p,\bar{p}\rangle~,\cr
\partial_{-}\langle U(x)|l,p,\bar{p}\rangle &= i\cos(2r)\langle U(x)|L_3|l,p,\bar{p}\rangle+  i\sin(2r)\cos(\theta_+)\langle U(x)|L_1|l,p,\bar{p}\rangle \cr
&\hspace{0.4cm} - i\sin(2r)\sin(\theta_+) \langle U(x)|L_2|l,p,\bar{p}\rangle~,\cr
\partial_r \langle U(x)|l,p,\bar{p}\rangle &= 2i\cos(\theta_+)\langle U(x)|L_2|l,p,\bar{p}\rangle+2i\sin(\theta_+)\langle U(x)|L_1|l,p,\bar{p}\rangle~,\label{eq:dm}
\end{align}
where we introduced the coordinates  
\be
\theta_{\pm} =\phi \pm \tau~,\qquad \partial_{\pm}= {\partial \over \partial \theta_\pm}~.
\ee
Inverting the relationship between $\partial_a \langle U(x)|l,p,\bar{p}\rangle$ and $\langle U(x)|L_a|l,p,\bar{p}\rangle$ leads to
\begin{align}\label{eq:zeta}
\zeta_1 &= -i\frac{\cos\theta_{+}}{\sin{(2r)}} \left(\partial_{-} + \cos{(2r)}\, \partial_{+}\right) - \frac{i}{2} \sin\theta_{+} \partial_r~,\cr
\zeta_2 &= i \frac{\sin\theta_{+}}{\sin{(2r)}}\left(\partial_{-} + \cos{(2r)}\, \partial_{+}\right) - \frac{i}{2} \cos\theta_{+} \partial_r ~,\cr
\zeta_3 &= i \partial_{+}~,
\end{align}
or, in terms of $\zeta_\pm = \zeta_1 \pm i \zeta_2$,
\begin{align}
\zeta_\pm = - i e^{\mp i \theta_+}\left(\csc{(2r)} \,\partial_{-} + \cot{(2r)}\, \partial_{+} \right)\pm \frac{1}{2} e^{\mp i \theta_+} \partial_r~,
\end{align}
and $\zeta_0=\zeta_3$.
These are simply three of the Killing vectors for $S^3$, which together satisfy one copy of the $su(2)$ algebra. 

To do the equivalent calculation for the barred sector, we should instead write
\begin{align}
|U(x)\rangle&=G(g^{-1}_L(x))\bar{G}(g_R(x))|\Sigma\rangle \nonumber \\
&=\bar{G}\left[g_R(x)\Sigma^{-1}g_L(x)\Sigma\right]|\Sigma\rangle \nonumber\\
&=\bar{G}\left[\Sigma^{-1}\widetilde{g}_R(x) g_L(x) \Sigma\right]|\Sigma\rangle \nonumber \\
&=\Sigma^{-1}e^{i\theta_{-}\bar{L}_3}e^{-2ir\bar{L}_2}e^{-i\theta_{+}\bar{L}_3}\Sigma|\Sigma\rangle~.\label{Uket}
\end{align}
This, after all, is the purpose of our definition of $\Sigma$: it lets us intertwine the two copies of $su(2)$. Therefore, the exact action of $\Sigma$ on group elements will affect the result of this calculation. We have two choices of $\Sigma$, given in \eqref{eq:Ishdef} and \eqref{eq:crossdef}, 
\begin{align}
\Sigma_{\text{cross}}=e^{i\pi \bar{L}_1} ~,\qquad 
\Sigma_{\text{Ish}}=e^{i\pi \bar{L}_2}~.
\end{align}
Working out the effect of the Ishibashi state in \eqref{Uket} we find
\begin{equation}
\Sigma_{\rm Ish}^{-1}e^{i\theta_{-}\bar{L}_3}e^{-2ir\bar{L}_2}e^{-i\theta_{+}\bar{L}_3}\Sigma_{\rm Ish}=e^{-i\theta_{-}\bar{L}_3}e^{-2ir\bar{L}_2}e^{i\theta_{+}\bar{L}_3}~,
\end{equation}
in other words conjugation by $\Sigma_{\rm Ish}$ flips $\theta_\pm \rightarrow - \theta_\pm$ while leaving $r$ fixed. For the crosscap state we instead find
\begin{equation}
\Sigma_{\rm cross}^{-1}e^{i\theta_{-}\bar{L}_3}e^{-2ir\bar{L}_2}e^{-i\theta_{+}\bar{L}_3}\Sigma_{\rm cross}=e^{-i\theta_{-}\bar{L}_3}e^{2ir\bar{L}_2}e^{i\theta_{+}\bar{L}_3}~,
\end{equation}
so that conjugation by $\Sigma_{\rm cross}$ flips $\theta_\pm \rightarrow - \theta_\pm$ and in addition $r\rightarrow -r$. 
From here on, the calculation to build $\bar \zeta_a$ is very similar to the unbarred case, but there will be differences depending on the choice of $\Sigma$. First, solving \eqref{eq:diffopLbar}, for $\Sigma= \Sigma_{\rm Ish}$ we find
\begin{align}\label{eq:barzeta}
\bar{\zeta}_1 &= - i \frac{\cos{\theta_{-}}}{\sin{2r}}\left(\partial_{+} + \cos{2r} \partial_{-}\right) -\frac{i}{2}\sin{\theta_{-}}\partial_r~,\cr
\bar{\zeta}_2 &= - i \frac{\sin{\theta_{-}}}{\sin{2r}}\left(\partial_{+} + \cos{2r} \partial_{-}\right) +\frac{i}{2}\cos{\theta_{-}}\partial_r~,\cr
\bar{\zeta}_3 &= -i \partial_{-}~,
\end{align}
or in terms of $\bar{\zeta_\pm} = \bar{\zeta_1} \pm i \bar{\zeta_2}$,
\be 
\bar{\zeta}_\pm =  - i e^{\pm i \theta_{-}}\left(\csc{2r} \partial_{+} + \cot{2r} \partial_{-}\right) \mp \frac{1}{2} e^{\pm i \theta_-} \partial_r~,
\ee
and $\bar\zeta_0=\bar \zeta_3$. These are the three additional Killing vectors for $S^3$, which are related to \eqref{eq:zeta} by the replacement $\theta_{\pm} \rightarrow -\theta_{\mp}$ and $r\rightarrow-r$. Together the generators $ \zeta_a$ satisfy the $su(2)_L$ algebra, while $\bar \zeta_a$ correspond to the generators of the second $su(2)_R$. Selecting $\Sigma=\Sigma_{\rm cross}$ is not dramatically different: we will again obtain \eqref{eq:barzeta} with $r\to -r$, and that flips the overall sign in $\bar \zeta_{1,2}$. Hence we will again find the second  copy of Killing vectors obeying $su(2)_R$; the difference at this stage between the two singlet states is an orientation of $r$ that does not affect the interpretation of $(\zeta_a,\bar\zeta_a)$ as the six Killing vectors for $S^3$.

Now we would like to find explicit expressions for  $\Phi_{p,\bar p}$. The procedure for either $\Sigma_{\rm Ish}$ or $\Sigma_{\rm cross}$ would produce the same special functions, with the difference being an overall normalization that depends on $(p,\bar p)$. For concreteness we will just focus on $\Sigma_{\rm Ish}$.   

We can construct the pseudofields by first solving for a highest weight state $\Phi_{0,0}$, and then acting with $(\zeta_{-})^p$ and $(\bar{\zeta}_{-})^{\bar p}$ on this solution to generate $\Phi_{p,\bar{p}}$. This will give a position-space representation of our abstract states $|l,p,\bar{p}\rangle$. The highest weight state satisfies
\begin{alignat}{2}
\zeta_3 \Phi_{0,0}&=\bar{\zeta}_3 \Phi_{0,0}&&=l\Phi_{0,0}~,\\
\zeta_{+}\Phi_{0,0}&=\bar{\zeta}_+ \Phi_{0,0}&&=0~.
\end{alignat}
These equations are solved by 
\begin{equation}
\Phi_{0,0}(r,\tau,\phi) = \langle U(x)|l,0,0\rangle =e^{-2il\tau}\cos^{2l}(r)~.
\end{equation}
The descendant states are then given by
\begin{align}\label{eq:basisphipp}
\Phi_{p,\bar{p}}(r,\tau,\phi)&=c_{p\bar{p}} e^{-2il\tau} \cos^{2l}(r) e^{i(p \theta_{+}-\bar{p} \theta_{-})} \tan^{\bar{p}-p}(r) P_{p}^{\bar{p}-p,-(2l+1)}\left(1+2\tan^2(r)\right)~,~ \nonumber \\
c_{p \bar{p}}&=(-1)^p\sqrt{\frac{p!(\bar{p}-(2l+1))!}{\bar{p}!(p-(2l+1))!}}~,
\end{align}
where here $P_n^{\alpha,\beta}(x)$ is a Jacobi polynomial. These satisfy \eqref{eq:Lplusop}-\eqref{eq:L0op} and their barred analogues.

\subsubsection{Wavefunction for the singlet states}

Where does our singlet state $|\Sigma\rangle$ sit on $S^3$? This question is ambiguous, since the answer depends on a choice of gauge. In the context of the discussion presented here, positions will depend  on how one selects the midpoint in \eqref{eq:localchoice}. Still it is instructive to answer it for the simple purpose of illustrating what our prior choices imply.  

Consider first the Ishibashi state $|\Sigma_{\rm Ish}\rangle$. To see the position of this state in $S^3$, it is very clear that at $r=0$, we have
\begin{equation}
\Phi_{p, \bar{p}}(\tau,r=0,\phi)=(-1)^pe^{-2i\tau(l-p)} \delta_{p,\bar{p}}~,
\end{equation}
which follows from \eqref{eq:basisphipp}. 
This is to be expected since $p \neq \bar{p}$ introduces a $\phi$ dependence which we know is absent at $r=0$.
Therefore, we can write
\begin{align}\label{eq:uish11}
|U(\tau,r=0)\rangle=\sum_{p} (-1)^p e^{2i\tau(l-p)} |l,p,p\rangle~,
\end{align}
which at $\tau=0$ is simply the Ishibashi state \eqref{eq:sish}. Thus we see that our Ishibashi state lives at $(r=0,\tau=0)$. If we had constructed a basis of $\Phi_{p,\bar{p}}$ from the $(\zeta_a,\bar{\zeta}_a)$ obtained from the crosscap states rather than the Ishibashi states, we would have seen that the crosscap state sits at $(r=0,\tau=0)$.

The wave function we would attribute to the Ishibashi state can also be explicitly calculated:
\begin{align}
\langle \Sigma_{\text{Ish}}|U(x)\rangle&=\frac{\left(\cos(\Theta_{\rm NPole})-i\sin(\Theta_{\rm NPole})\right)^{2l+1}}{2i\sin(\Theta_{\rm NPole})} \nonumber \\
&=\frac{e^{-2il\Theta_{\rm NPole}}}{1-e^{2i\Theta_{\rm NPole}}}~.
\end{align}
where $\Theta_{\rm NPole}$ is the geodesic distance \eqref{eq:distace} between $x$ and $r'=0,\tau'=0$ ---the North Pole of the three-sphere.\footnote{This corresponds to the North Pole of the $S^2$ time slices for Euclidean time $\tau'=0$. It is a point on Penrose diagrams, not a line.}

Still we stress that the values of $\tau$ and $r$ are somewhat artificial. For instance, in (\ref{eq:uish11}) the crosscap state can be seen to be related to the Ishibashi state by a simple shift in $\tau$.  This is a reflection of the fact that there is considerable gauge freedom in how we describe solutions.

\subsubsection{Wick rotation and quasi-normal modes}\label{sec:wick}

Before proceeding to discuss $SL(2,\mathbb{C})$ Chern-Simons theory, i.e. the Lorentzian formulation of dS$_3$ gravity, it is instructive to interpret our Euclidean results in Lorentzian signature.  We will simply now use a Wick rotation of the metric formulation to provide a first interpretation of  our results. As described in App.~\ref{app:ds3}, the metric analytic continuation is implemented by taking 
\be
t\rightarrow -i \ell \tau~.
\ee
The Wick-rotated $\Phi_{p,\bar{p}}$ in \eqref{eq:basisphipp} are therefore 
\begin{align}
\Phi_{p,\bar{p}}(r,t,\phi)&=c_{p\bar{p}} e^{il(\bar{z}-z)} \cos^{2l}(r) e^{i(p z-\bar{p} \bar z)} \tan^{\bar{p}-p}(r) P_{p}^{\bar{p}-p,-(2l+1)}\left(1+2\tan^2(r)\right)~,~ \nonumber \\
c_{p \bar{p}}&=(-1)^p\sqrt{\frac{p!(\bar{p}-(2l+1))!}{\bar{p}!(p-(2l+1))!}}~,
\end{align}
with $z\equiv\phi+it$, and $\bar{z}\equiv\phi-it$.
In terms of the more familiar hypergeometric functions and radial coordinate $u\equiv \sin(r)$, we have (using that $\Phi_{p,\bar{p}}=e^{2i(p-\bar{p})\phi} \Phi_{\bar{p},p}$):
\begin{align}
\Phi_{\omega,k}(r,t,\phi)&=c_{p\bar{p}} {\frac{\omega+|k|}{2}+l\choose \frac{\omega-|k|}{2}+l} (1-u^2)^{-\omega/2} u^{|k|} e^{-i k \phi} e^{-\omega t} { }_2F_1\left(\frac{|k|-\omega}{2}-l,\frac{|k|-\omega}{2}+l+1;|k|+1;u^2\right)~, \nonumber\\
\omega&=p+\bar{p}-2l>0 ~~,~~ k=\bar{p}-p~.
\label{eq:quasinormal}
\end{align}
Note that instead of oscillating in time, these functions are now purely decaying. In fact, the $\Phi_{\omega, k}$ are exactly (up to normalization) the quasi-normal modes of dS$_3$\cite{LopezOrtega:2006my, Anninos:2011af}. As discussed in Sec.\,\ref{sec:wilsongreen}, given a scalar field of mass $m$, there are two representations $\mathscr{R}_\pm$ that have the same Casimir: one with $l=-h$ and one with $l=h-1$ . These two representations have different characters (and thus Wilson lines), and both are needed to obtain the full Green's function: $G(\Theta)=a_h W_{\mathscr{R}_+}+a_{1-h}W_{\mathscr{R}_-}$. Each choice of $l$ matches one of the two distinct sequences of quasi-normal modes in dS$_3$. This reinforces the idea that both representations are needed to describe a bulk scalar field.

The Wick rotation can also be used to simply obtain Lorentzian Killing vectors from \eqref{eq:zeta} and \eqref{eq:barzeta}. These can then be re-organized in an $sl(2,\mathbb{C})$ representation in the following way: 
\begin{align}\label{eq:Wickrotation}
-i\zeta_1&~ \underset{\tau\to i t/\ell }{\longrightarrow}\quad\mathcal H_1 =-\frac{\cos{ z}}{\sin{2r}}\left(\bar \partial + \cos{2r} \partial\right) -\frac{1}{2}\sin{ z}\,\partial_r~,\nonumber\\
-i\zeta_2&~ \underset{\tau\to i t/\ell }{\longrightarrow}\quad\mathcal H_2 =\frac{\sin{ z}}{\sin{2r}}\left(\bar \partial + \cos{2r} \partial\right) -\frac{1}{2}\cos{ z}\,\partial_r~ \nonumber \\
\zeta_3& ~\underset{\tau\to i t/\ell }{\longrightarrow}\quad \mathcal H_3 = i\partial ~,~\nonumber \\
i\bar \zeta_1& ~\underset{\tau\to i t/\ell }{\longrightarrow}\quad\bar{\mathcal H}_1 =\frac{\cos \bar z}{\sin{2r}} \left(\partial + \cos{2r} \bar \partial\right) + \frac{1}{2} \sin \bar z \,\partial_r~,\nonumber\\
-i\bar \zeta_2& ~\underset{\tau\to i t/\ell }{\longrightarrow}\quad\bar{\mathcal H}_2 =- \frac{\sin \bar z}{\sin{2r}}\left(\partial + \cos{2r} \bar\partial\right) + \frac{1}{2} \cos \bar z \,\partial_r~,\nonumber\\
-\bar \zeta_3&~ \underset{\tau\to i t/\ell }{\longrightarrow}\quad\bar{\mathcal H}_3 = i\bar{\partial}~.
\end{align}
The operators $(\mathcal{H}_a,\bar{\mathcal{H}}_a)$ have been normalised such that they form an $sl(2,\mathbb{R})\times sl(2,\mathbb{R})$ algebra. More importantly,  these operators have a simple action on the quasinormal modes. We can see this explicitly by reorganizing the operators into the combinations
\begin{alignat}{2}
\mathcal H_0 &= -\mathcal H_3~,\indent &&\mathcal H_\pm =  \mathcal{H}_2 \mp i \mathcal{H}_1~, \nonumber \\
\bar{\mathcal H}_0 &= \bar{\mathcal H}_3~, \indent &&\bar{\mathcal{H}}_\pm = \bar{\mathcal{H}}_2 \pm i \bar{\mathcal{H}}_1~.
\end{alignat}
The quasinormal mode $\Phi_{00}$ is a highest weight state of our representation,
\be \mathcal H_+ \Phi_{00} = 0~,\ee
while the rest of the quasinormal modes obey
\begin{align}\label{eq:sl21}
\mathcal{H}_0 \Phi_{p \bar p} &=(h+p)\Phi_{p, \bar p}~, \nonumber\\
\mathcal{H}_{+}\Phi_{p \bar p}&=\sqrt{p(p+2h+1)}\Phi_{p-1, \bar p}~, \nonumber \\
\mathcal{H}_{-}\Phi_{p \bar p}&=\sqrt{(p+1)(p+2h)}\Phi_{p+1, \bar p}~,
\end{align}
and similarly for the barred sector. In this expression we have $h =-l$,\footnote{We are focusing here on $\mathscr{R}_+$ for notational simplicity. Analogous results with $h\rightarrow 1-h$ can be obtained for $\mathscr{R}_-$, which has $l=h-1$.} and hence the modes $\Phi_{p, \bar p}$  characterize a highest weight representations of $sl(2)$ with Casimir $h(h-1)$.  Furthermore, the (anti-)Hermitian properties of the $su(2)$ generators $L_{0,\pm}$ in \eqref{eq:Hermiticity} combined with the map
in \eqref{eq:Wickrotation}, dictate that the generators $\mathcal{H}_{0,\pm}$ have the usual Hermiticity properties. This makes the representations unitary when organized in terms of the $sl(2,\mathbb{R})$ basis. 

The Wick rotation gives an interpretation for the algebraic structure of the quasi-normal mode spectrum of the static patch. Our construction resonates with~\cite{Anninos:2011af}, where it was noticed that the quasinormal modes had a ``hidden'' $SL(2,\mathbb{R})$ symmetry, but the origin of this remained mysterious.  A similar result was found in \cite{Chatterjee:2016ifv}.  


Finally, the quasinormal modes additionally satisfy the Casimir equation for our representations,
\begin{equation}\label{eq:KGphi}
\left(\nabla^2+\bar\nabla^2\right) \Phi_{p,\bar{p}}(x)=2h(h-1)\Phi_{p,\bar{p}}(x)~.
\end{equation}
where $\nabla^2 = -\eta^{ab}{\cal H}_a{\cal H}_b$, and $\bar\nabla^2 = -\eta^{ab}\bar {\cal H}_a\bar {\cal H}_b$, so that $\nabla^2+\bar{\nabla}^2=-\frac{1}{2} \nabla^2_{dS_3}$ is the d'Alembertian on Lorentzian dS$_3$. With the insight of the Wick rotation, the representation \eqref{eq:sl21} will be our focus in the subsequent section as we study $SL(2,\mathbb{C})$ Chern-Simons theory.

\section{Wilson lines in $SL(2,\mathbb{C})$ Chern-Simons}\label{sec:analyticcont}

Everything we have discussed so far has been based on Euclidean dS$_3$. In this section, we discuss how our construction can be translated to Lorentzian signature, guided by the properties of our representation under analytic continuation. Based on the Euclidean analysis, we will select a suitable representation of $sl(2,\mathbb{C})$, and implement this choice for the inflationary patch of dS$_3$.

\subsection{Chern-Simons formulation of Lorentzian dS$_3$ gravity}
We start from $SL(2,\mathbb{C})$ Chern-Simons theory with action
\be S_{CS}[\mathcal A] = \frac{is}{4\pi} \int \mbox{Tr}\left(\mathcal A \wedge d\mathcal A + \frac{2}{3} \mathcal A\wedge \mathcal A \wedge \mathcal A\right)- \frac{i s}{4\pi} \int \mbox{Tr}\left(\bar{\mathcal A} \wedge d\bar{\mathcal A} + \frac{2}{3} \bar{\mathcal A}\wedge \bar{\mathcal A} \wedge \bar{\mathcal A}\right)~,\label{eq:CSactionLor}\ee
with ${\mathcal A}, \bar{\mathcal A} \in sl(2,\mathbb{C})$, and complex parameter $s$. The relation of \eqref{eq:CSactionLor} to Lorentzian dS$_3$ gravity was done in \cite{Witten:1989ip}, and more recent discussions include \cite{Fjelstad:2002wf,Govindarajan:2002ry,Balasubramanian:2002zh,Cotler:2019nbi}. To build this gravitational interpretation, we expand the gauge fields over the generators $\mathcal L_a~, \bar{\mathcal{L}}_a$  of $sl(2,\mathbb{C})$ as
\begin{align}
\mathcal A = -\left(i\omega^a +\frac{1}{\ell}e^a\right)\mathcal L_a~, \indent \bar{\mathcal A} =  -\left(i\omega^a -\frac{1}{\ell}e^a\right)\bar{\mathcal L}_a~.\label{eq:LorA}
\end{align}
where the $sl(2,\mathbb{C})$ generators can be related to the generators of $so(1,3)$ isometries as
\begin{align}
\mathcal L_a = \frac{i}{2}(\mathcal J_a + i \ell \mathcal P_a)~, \indent \bar{\mathcal L}_a = \frac{i}{2}(\mathcal J_a - i \ell \mathcal P_a)~.\label{eq:L}
\end{align}
They satisfy the algebra
\begin{align} \label{eq:copiessl2c}
[\mathcal L_a, \mathcal L_b] &= i \epsilon_{abc} \mathcal L^c~, \cr
[\bar{\mathcal L}_a, \bar{\mathcal L}_b] &= i \epsilon_{abc} \bar{\mathcal L}^c~,\cr
[\mathcal L_a, \bar{\mathcal L}_b] &= 0~,
\end{align}
with indices raised by $\eta^{ab}$, and we take the convention that $\eta^{11}=\eta^{22}=+1$ and $\eta^{33}=-1$. The trace is taken with the bilinear form
\be \label{eq:tracesl2c}
\mbox{Tr}(\mathcal L_a \mathcal L_b) = \mbox{Tr}(\bar{\mathcal L}_a \bar{\mathcal L}_b) = -\frac{1}{2} \eta_{ab}~.
\ee
Using \eqref{eq:LorA}, the action \eqref{eq:CSactionLor} becomes
\begin{equation}
S_{EH}=\frac{s}{2\pi\ell} \int_\mathcal{M} \left[e^a\wedge\left(d\omega_a+\frac{1}{2}\epsilon_{abc} \omega^b \wedge \omega^c\right)-\frac{1}{6\ell^2} \epsilon_{abc} e^a \wedge e^b \wedge e^c\right]~.
\end{equation}
This reduces to the Einstein-Hilbert action with positive cosmological constant given the identification
\be 
s = \frac{\ell}{4G_3} ~ \in \mathbb{R}~.
\ee

It is important to note that $\mathcal A$ and $\bar{\mathcal A}$ are not independent variables. They are related by complex conjugation, and this relation depends on how we choose to relate ${\mathcal L}_a$ to $\bar{\mathcal L}_a$. For now it suffices to demand \eqref{eq:tracesl2c}, which assures reality of the action \eqref{eq:CSactionLor}, and we will constrain further the representation as we construct the appropriate probes.

\subsection{Construction of probes in $sl(2,\mathbb{C})$}

As in the Euclidean case, we would like to build probes in $SL(2, \mathbb{C})$ Chern-Simons theory via the Wilson line operator \eqref{eq:wilsonline}. 
The most natural choice is to simply implement the discrete highest weight representation we inferred in Sec.\,\ref{sec:wick} from the Euclidean theory. For a further motivation of this choice using an analytic continuation of the $SO(4)$ and $SL(2,\mathbb{C})$ Chern-Simons theories, see App.~\ref{app:cont}. In the language of the $SL(2,\mathbb{C})$ Chern-Simons, we will build this representation by using the $sl(2)$ generators\footnote{A similar discussion regarding representations of $sl(2,\mathbb{C})$ is discussed in \cite{Chatterjee:2016ifv}. One difference is that the authors take $sl(2,\mathbb{C})\sim su(1,1)\times su(1,1)$. }
\begin{align}
\mathcal{L}_0=-\mathcal{L}_3~, \qquad
\mathcal{L}_\pm = \mathcal{L}_2 \mp i \mathcal{L}_1~,\label{Eq.Lpmdef}
\end{align}
with algebra
\begin{equation}
[\mathcal{L}_0,\mathcal{L}_\pm]=\mp \mathcal{L}_\pm~~,~~[\mathcal{L}_+,\mathcal{L}_-]=2\mathcal{L}_0~.
\end{equation}
The highest weight representation in this basis satisfies
\begin{align}\label{eq:unitaryrep}
\mathcal{L}_0|h,p\rangle&=(h+p)|h,p\rangle~, \nonumber\\
\mathcal{L}_{+}|h,p\rangle&=\sqrt{p(p+2h+1)}|h,p-1\rangle ~,\nonumber \\
\mathcal{L}_{-}|h,p\rangle&=\sqrt{(p+1)(p+2h)}|h,p+1\rangle~,
\end{align}
where $p$ is a positive integer. For now, we take $h$ to be a real parameter that controls the Casimir of the representation
\begin{align}
-\eta^{ab}\mathcal{L}_a \mathcal{L}_b|h,p\rangle&=( \mathcal{L}_0^2- \mathcal{L}_+\mathcal{L}_- -\mathcal{L}_-\mathcal{L}_+)|h,p\rangle\cr& = h(h-1)|h,p\rangle~.
\end{align}
Of course, we anticipate that this parameter will match $h=\frac{1+\sqrt{1-(m\ell)^2}}{2}$ (or the other solution which gives the same Casimir).
In addition we demand the operators satisfy $\mathcal{L}_0^\dagger= \mathcal{L}_0$ and $\mathcal{L}_{\pm}^\dagger = \mathcal{L}_{\mp}$;  this makes the representation unitary.
 For the barred sector we also select a highest-weight representation of $sl(2,\mathbb{R})$, which obeys
\begin{align}\label{eq:unitaryrepbar}
\bar{\mathcal{L}}_0|\bar h,\bar{p}\rangle&=(\bar h+\bar p)|\bar h,\bar p\rangle ~,\nonumber\\
\bar{\mathcal{L}}_{+}|\bar h,\bar{p}\rangle&=\sqrt{\bar p(\bar p+2\bar h+1)}|\bar h,\bar p-1\rangle~, \nonumber \\
\bar{\mathcal{L}}_{-}|\bar h,\bar{p}\rangle&=\sqrt{(\bar p+1)(\bar p+2\bar h)}|\bar h,\bar p+1\rangle~.
\end{align}
The quadratic Casimir for this sector is
\begin{align}
-\eta^{ab}\bar{\mathcal{L}}_a \bar{\mathcal{L}}_b|\bar h,\bar{p}\rangle&= 2\bar h(\bar h-1)|\bar h,\bar{p}\rangle~. 
\end{align}

Singlet states in this case are defined in an analogous way as in Sec.\,\ref{sec:singlet}: we will consider two possible conditions
\begin{align}
({\cal L}_ k- (-1)^k \bar {\cal L}_{-k}) \left|\Sigma_{\rm cross}\right> = 0~,\cr
({\cal L}_k - \bar{\cal L}_{-k}) \left|\Sigma_{\rm Ish}\right>=0~,
\end{align}
for $k=0,\pm$, and the solutions are
\begin{align}\label{eq:singletsl2c}
\left|\Sigma_{\rm Ish}\right> &= \sum_{p=0}^\infty \left|h,p,p\right>~,\cr
\left|\Sigma_{\rm cross}\right> &= \sum_{p=0}^\infty (-1)^p \left|h,p,p\right>~,
\end{align}
where the singlet condition sets $h=\bar h$, and we are using $\left|h,p,\bar p\right> \equiv \left|h,p\right> \otimes \left|h,\bar p\right>$. There is a difference in that the $(-1)^p$ factor appears for the crosscap state rather than for Ishibashi. This results from the fact that \eqref{eq:unitaryrep} and \eqref{eq:unitaryrepbar} do \emph{not} contain a minus sign. In this sense they more closely resemble the AdS$_3$ rather than EdS$_3$ versions.

There is, however, a more important conceptual difference when we move to Lorentzian de Sitter. Recall that in EdS$_3$ the singlet states played a role in relating the two (barred and unbarred) copies of $SU(2)$, which are initially independent; in the same way, here they allow us to relate two copies of $SL(2,\mathbb{R})$. Since in $SL(2,\mathbb{C})$ Chern-Simons theory the components $\mathcal A_a$ and $\bar{\mathcal A}_a$ are related by complex conjugation to ensure the reality of the Einstein-Hilbert action, the choice of a singlet state additionally picks out a reality condition on the fields propagating on the background created by $\mathcal A$ and $\bar{\mathcal A}$.

We can now evaluate the Wilson line. We are treating $sl(2,\mathbb{C})$ as two copies of $sl(2)$, as decomposed in \eqref{eq:copiessl2c}, and hence we want to evaluate 
\begin{align}
W_{\mathscr{R}}(x_i, x_f)=  \left<\Sigma \right| \mathscr{P}\mbox{exp}\left(-\int_\gamma {\cal A}\right)\mathscr{P}\mbox{exp}\left(-\int_\gamma \bar {\cal A}\right)\left|\Sigma\right>~,
\end{align}
where we selected the endpoint states to be one of the singlet states in \eqref{eq:singletsl2c}$: |U_{i,f}\rangle =|\Sigma\rangle$. Writing this as group elements acting on each copy of $sl(2)$ we have
\begin{align}\label{eq:derivationXX}
W_{\mathscr{R}}(x_i, x_f) &= \langle \Sigma |G({\cal L}) \bar G({\cal R}^{-1}) \left|\Sigma\right>\cr
&= \langle \Sigma |G({\cal L} \, \tilde {\cal R}) \left|\Sigma\right>\cr
&= \sum_{p=0}^\infty \langle h,p|G( {\cal L} \, \tilde {\cal R})|h,p\rangle\cr
&= \frac{e^{i h \alpha}}{1-e^{i\alpha}}~,
\end{align}
where 
\begin{align}
{\cal A}&= g \dd g^{-1}~,\qquad {\cal L}\equiv g(x_f)g(x_i)^{-1}~, \cr
\bar {\cal A} &= \bar g^{-1}\dd \bar g~,\qquad  {\cal R}^{-1}\equiv \bar g(x_f)^{-1} \bar g(x_i) ~,
\end{align}
and $\tilde {\cal R} = \Sigma\, {\cal R}\, \Sigma^{-1}$. As before, we have defined $\alpha$ by assuming we can diagonalize the group element as
\begin{equation}\label{eq:diagonalization}
{\cal L}\, \Sigma \, {\cal R} \, \Sigma^{-1}=V^{-1} e^{i\alpha \mathcal L_0} V~.
\end{equation}
Other than the fact that we are using the states $|h, p, \bar p \rangle$ and generators $\mathcal L_a$ associated to our unitary Lorentzian representation rather than the states $|l, p, \bar p\rangle$ and generators $L_a$ for the non-unitary Euclidean representation, everything proceeds as for the Euclidean case. In the end we can recognize that the Lorentzian Wilson line is just a character associated to our Lorentzian representations.

\subsection{Inflationary patch}

In this final portion we will consider the inflationary patch of dS$_3$ in order to illustrate our Lorentzian construction. The line element reads
\be\label{eq:inflation}
 \frac{\dd s^2}{\ell^2} = \frac{1}{\eta^2}\left(-\dd\eta^2 +  \dd w \dd\bar w\right)~,
\ee
where $\eta>0$, positive timelike infinity is located at $\eta\to0$, and $w=x+i y$ is a complex variable. See App.~\ref{app:coord} for a review of these coordinates.

For the inflationary patch, we use the group elements
\begin{align}
g= e^{-\frac{i w}{\eta}\mathcal L_+ } e^{\log{\eta} \,\mathcal L_0}~,\qquad
\tilde {\bar g} = e^{\log{\eta} \,\mathcal L_0} e^{\frac{i \bar w}{\eta}\mathcal L_- }~.\label{eq:ginfl}
\end{align}
These give connections
\be\label{eq:connection1}
{\mathcal A} =g \dd g^{-1}= -\frac{\dd \eta}{\eta} \mathcal L_0 + \frac{i \dd w}{\eta} \mathcal L_+~, \indent
\bar{\mathcal A} =\tilde {\bar g}^{-1}\dd \tilde {\bar g}=  \frac{\dd \eta}{\eta} \mathcal L_0 + \frac{i \dd\bar{w}}{\eta} \mathcal L_- ~.
\ee
In our conventions the Lorentzian metric is 
\be
g_{\mu \nu} =- {\ell^2\over 2} {\rm Tr}\le[ \le( {\mathcal A}_\mu -\bar{\mathcal A}_\mu\ri)\le( {\mathcal A}_\nu - \bar{\mathcal A}_\nu\ri)\ri]~,
\ee
where here we are using the same generators for barred and unbarred connections. It is easy to check this reproduces \eqref{eq:inflation}. 

As in the Euclidean case, we can define the local state from the group elements acting on the singlet state,
\begin{align} 
|U(x)\rangle&=G(g(x)^{-1}) \bar{G}(\bar{g}(x))|\Sigma\rangle~,\cr
&=G(g(x)^{-1} \tilde {\bar g}(x)^{-1})|\Sigma\rangle~,
\end{align}
where $\tilde {\bar g}= \Sigma\,  {\bar g}\,\Sigma^{-1}$. Evaluating this using the group elements \eqref{eq:ginfl}, we find
\begin{align}
|U(x)\rangle =e^{-\log{\eta} \,\mathcal L_0} e^{\frac{i w}{\eta}\mathcal L_+ } e^{-\frac{i \bar w}{\eta}\mathcal L_- }e^{-\log{\eta} \,\mathcal L_0} |\Sigma\rangle ~.
\end{align}

Now we will construct local pseudofields from the states $|U(x)\rangle$. We follow an exactly analagous procedure to the EdS$_3$ case in Sec.\,\ref{sec:localbasis}, starting with expansion of the state over the states $|h,p,\bar p\rangle$ that form a basis for our unitary Lorentzian representations,
\be |U(x)\rangle = \sum_{p,\bar p=0}^\infty \Phi^*_{p, \bar p}(x) |h, p, \bar p\rangle~.\ee
Inverting this relation gives
\begin{align}
\Phi_{p, \bar p}(x) &= \langle U(x)| h, p, \bar p\rangle~.
\end{align} 

We can define a set of differential operators ${\cal H}_a$ and $\bar{\cal H}_a$ as
\begin{align} 
\langle U(x) | \mathcal L_a | h, p, \bar p\rangle =\mathcal{H}_a \langle U(x)| h, p, \bar p\rangle~, \label{eq:diffopL}\\
\langle U(x) | \bar{\mathcal L}_a | h, p, \bar p\rangle =  \bar{\mathcal H_a} \langle U(x)| h, p, \bar p\rangle~. \label{eq:diffopLbar}
\end{align}
Taking derivatives of the pseudofield $\Phi_{p,\bar{p}}(x)=\langle U(x)|h,p,\bar{p}\rangle$, we find
\begin{align}
\partial \langle U(x)|h,p,\bar{p}\rangle &= \frac{i}{\eta^2}\langle U(x)|\mathcal L_+|h,p,\bar{p}\rangle - \frac{i \bar{w}^2}{\eta^2}\langle U(x)|\mathcal L_-|h,p,\bar{p}\rangle+ \frac{2\bar{w}}{\eta^2}\langle U(x)|\mathcal L_0|h,p,\bar{p}\rangle~, \cr
\bar \partial \langle U(x)|h,p,\bar{p}\rangle &= -i \langle U(x)|\mathcal L_- |h,p,\bar{p}\rangle~,\cr
\partial_\eta \langle U(x)|h,p,\bar{p}\rangle &= \frac{2i \bar{w}}{\eta}\langle U(x)|\mathcal L_- |h,p,\bar{p}\rangle-\frac{2}{\eta}\langle U(x)|\mathcal L_0|h,p,\bar{p}\rangle~,
\end{align}
and from here we find
\begin{align}
\mathcal H_+ &= -i (\eta^2 \partial  + \eta \bar{w} \partial_\eta + \bar{w}^2 \bar{\partial})~,\cr
\mathcal H_- &= i \bar{\partial} ~,\cr
\mathcal H_0 &= - \frac{\eta}{2} \partial_\eta  - \bar{w} \bar{\partial} ~.
\end{align}
These are three Killing vectors for the inflationary patch of dS$_3$, whose boundary limits $\eta\rightarrow 0$ give one (barred) set of conformal generators.

The state $|U(x)\rangle$ can be equivalently be written in terms of the barred sector as
\begin{align} 
| U(x) \rangle &=\bar{G}\left[\Sigma^{-1}\widetilde{\bar g}(x) g(x) \Sigma\right]|\Sigma\rangle \nonumber \\
&= \Sigma^{-1} e^{\log{\eta} \,\bar{\mathcal L}_0} e^{\frac{i \bar w}{\eta}\bar{\mathcal L}_- } e^{-\frac{i w}{\eta}\bar{\mathcal L}_+ } e^{\log{\eta} \,\bar{\mathcal L}_0} \Sigma |\Sigma \rangle~,
\end{align}
where we have initially kept the state $\Sigma$ arbitrary. Using the definitions \eqref{eq:Ishdef} and \eqref{eq:crossdef} for the Ishibashi and crosscap states through their action on generators, for the Ishibashi state conjugation gives
\be  \Sigma_{\rm Ish}^{-1} \,e^{\log{\eta} \,\bar{\mathcal L}_0} e^{\frac{i \bar w}{\eta}\bar{\mathcal L}_- } e^{-\frac{i w}{\eta}\bar{\mathcal L}_+ } e^{\log{\eta} \,\bar{\mathcal L}_0}\, \Sigma_{\rm Ish} = e^{-\log{\eta} \,\bar{\mathcal L}_0} e^{-\frac{i \bar w}{\eta}\bar{\mathcal L}_+ } e^{\frac{i w}{\eta}\bar{\mathcal L}_- } e^{-\log{\eta} \,\bar{\mathcal L}_0}~, \ee
while for the crosscap state,
\be  \Sigma_{\rm cross}^{-1}\, e^{\log{\eta} \,\bar{\mathcal L}_0} e^{\frac{i \bar w}{\eta}\bar{\mathcal L}_- } e^{-\frac{i w}{\eta}\bar{\mathcal L}_+ } e^{\log{\eta} \,\bar{\mathcal L}_0}\, \Sigma_{\rm cross} = e^{-\log{\eta} \,\bar{\mathcal L}_0} e^{\frac{i \bar w}{\eta}\bar{\mathcal L}_+ } e^{-\frac{i w}{\eta}\bar{\mathcal L}_- } e^{-\log{\eta} \,\bar{\mathcal L}_0}~. \ee
Restricting to the Ishibashi state for definiteness, we can follow a similar procedure and solve for the barred differential operators. We find
\begin{align}
\bar{\mathcal H}_+ &= i (\eta^2 \bar \partial  + \eta w \partial_\eta  + w^2 \partial)~,\cr
\bar{\mathcal H}_- &= -i \partial ~,\cr
\bar{\mathcal H}_0 &= - \frac{\eta}{2} \partial_\eta  - w \partial ~.
\end{align}
Thus there is again a simple relation between the barred and unbarred differential operators. For the Ishibashi state the barred sector amounts to taking $w\leftrightarrow -\bar{w}$. The procedure can be repeated for the crosscap state, and in that case we must take $w\leftrightarrow \bar{w}$. We obtain from this a second set of Killing vectors whose $\eta\rightarrow0$ limit matches onto the second (unbarred) set of conformal generators.

Now we can build solutions that explicitly realize our unitary representations. The highest weight state satisfies
\begin{alignat}{2}
\mathcal{H}_0 \Phi_{0,0}&=\bar{\mathcal H}_0 \Phi_{0,0}&&=h\Phi_{0,0}~,\\
\mathcal{H}_{+}\Phi_{0,0}&=\bar{\mathcal H}_+ \Phi_{0,0}&&=0~,
\end{alignat}
and this equation is solved by
\begin{equation}
\Phi_{0,0}(\eta, w, \bar{w}) = \langle U(x)|h,0,0\rangle =\eta^{2h}(\eta^2-w \bar{w})^{-2h}~.
\end{equation}
We can again build the descendents by lowering starting from this highest weight state. For the case $p>\bar{p}$ we find
\be \Phi_{p, \bar p}(\eta,w, \bar w) =b_{p,\bar p}\,\eta^{2h+2n} w^{p-\bar p} (\eta^2-w \bar{w})^{-p-\bar p -2h} P_n^{(|p-\bar p|, -p -\bar{p} -2h)}\left(1-\frac{2w\bar w}{\eta^2}\right)~,\ee
where
\be b_{p,\bar p}= i^{p}(-i)^{\bar{p}} \sqrt{\frac{\bar{p}! (p+2h-1)!}{p!(\bar{p}+2h-1)!}} ~,\qquad n = \frac{1}{2}(p + \bar p - |p-\bar p|)~. \ee
For $\bar{p}>p$, the solution is $\Phi_{p, \bar p}(\eta,w,\bar w) = (-i)^p i^{\bar p}  \Phi_{\bar p, p}(\eta, \bar w, w)$. The solutions are again Jacobi polynomials $P_n^{\alpha, \beta}(x)$, however in this case $n$ depends nontrivially on both quantum numbers $p,\bar{p}$. Just like the static patch quasinormal modes, these are eigenfunctions satisfying \eqref{eq:sl21} and they solve the Klein-Gordon equation \eqref{eq:KGphi} in inflationary coordinates.

Restricting to $w=\bar{w}=0$ at finite $\eta$, the solution for the pseudofield reduces to
\be \Phi_{p, \bar p}(\eta, 0, 0) = \eta^{-2(p+h)} \delta_{p, \bar p}~.\ee
This means we can write
\be |U(\eta, 0, 0)\rangle = \sum_p \eta^{-2(p+h)} |h, p, p\rangle~, \label{eq:ish-infl}\ee
which at $\eta=1$ is simply the Ishibashi state, \eqref{eq:singletsl2c}. Thus we see that our Lorentzian Ishibashi state lives at $w=\bar{w}=0, \eta=1$. By going to embedding coordinates \eqref{eq:infl-coords}, it is easy to see that, up to analytic continuation, this is the same bulk point as $r=0, \tau=0$ where the Ishibashi state was located in static coordinates. Of course, once again we note that there is nothing special about that point: it is simply the product of various gauge choices we made along the way.

Finally we turn to the Wilson line, which can be evaluated directly as
\begin{align}  
W_{\mathscr{R}}(x_i, x_f) &= \langle \Sigma| G(g(x_f) g(x_i)^{-1}) \bar{G}(\tilde{\bar g}(x_f)^{-1} \tilde{\bar g}(x_i))| \Sigma \rangle\cr
&= \langle \Sigma|  G(g(x_f) g(x_i)^{-1} \tilde{\bar g}(x_i)^{-1} \tilde{\bar g}(x_f)) | \Sigma \rangle~.
\end{align}
Using \eqref{eq:diagonalization} and the explicit inflationary group elements \eqref{eq:ginfl}, we can solve for the parameter $\alpha$ describing the eigenvalue of the group element. We find
\be \cos{\left(\frac{\alpha}{2}\right)} = \frac{\eta_i^2 + \eta_f^2 - (w_f - w_i)(\bar{w}_f-\bar{w}_i)}{2\eta_i \eta_f}~.\ee
The right hand side is again just the invariant distance but now in inflationary coordinates (see App.~\ref{sect:prop}). This is directly analagous to our analysis of the Euclidean case, where $\alpha$ was related to the invariant distance in Hopf coordinates. We again have 
\be\label{eq:alphainfl}
\alpha=\pm 2\Theta + 4\pi n~, \qquad n\in \ZZ ~.
\ee
We can now relate the Wilson line to a Green's function. Recall that the Lorentzian Wilson line was equal to a character of our representation,
\be W_{\mathscr{R}}(x_i, x_f) = \frac{e^{i h \alpha}}{1-e^{i\alpha}}~.\ee
Using \eqref{eq:alphainfl}, we can convert this to a function of the invariant distance. After again defining
\be 
a_h=\frac{i}{2\pi \ell} \frac{1}{1-e^{-4\pi i h}}~,
\ee
we find that taking the irreducible representation $\mathscr{R}_+$ with $h=\frac{1+\sqrt{1-(m\ell)^2}}{2}$ leads to
\begin{equation}
W_{\mathscr{R}_+}(x_i,x_f)=\frac{1}{a_h} \mathcal{G}_h(\Theta)~.
\end{equation}
As in the Euclidean case given by \eqref{eq:eucgreen111}, to obtain the Green's function \eqref{eq:Greensfn} it is necessary to use both representations $\mathscr{R}_\pm$ with highest weight $h$ and $1-h$, 
\begin{equation}
G(\Theta)=a_h W_{\mathscr{R}_+}(x_i,x_f)+a_{1-h}W_{\mathscr{R}_-}(x_i,x_f)~.
\end{equation}

\section{Discussion}\label{sec:discussion}

In this last section we highlight our main findings and discuss some interesting future directions. 

\paragraph{Singlet states in 3D de Sitter.} 
To summarize: the singlet states we constructed in Sec.\,\ref{sec:wl} take the form
\begin{equation}
|\Sigma\rangle=\sum_{p,\bar p} a_{p,\bar p}|l,p,\bar p\rangle~,
\end{equation}
where $|l,p,\bar{p}\rangle = |l,p\rangle \otimes |l,\bar{p}\rangle$  are basis vectors of a non-unitary representation of $su(2)$. One of the consequences tied to selecting this unconventional representation is that we have a continuous parameter that we can identify with the mass of particle: we take $-1<l <0$, and its relation to the mass is $4l(l+1)=-{m^2 \ell^2}$. Although our discussion is limited to masses in the ranges $0<m^2\ell^2 <1$, our approach should be easily extendable to allow for arbitrary positive values of $m^2\ell^2$. Note that the representation we consider for Lorentzian de Sitter spacetime, which results from analytic continuation of the nonunitary $su(2)$ representation, is just the discrete series representation of $sl(2,\mathbb{R})$. We expect that extending the mass range would require building non-unitary representations of $su(2)$ that resemble the continuous series in $sl(2,\mathbb{R})$, which includes the principal and complementary series representations~\cite{Chatterjee:2016ifv}.

These singlet states are very reminiscent of the description of bulk local states in AdS. In \cite{Verlinde:2015qfa,Miyaji:2015fia,NakayamaOoguri2015}, it was shown that a  bulk field configuration at the centre of AdS corresponds to a crosscap state in the CFT. While there are certainly similarities between the two stories (emphasized by our choice of terminology for singlet states), there are also some notable differences. In the context of AdS/CFT, the crosscap states are states in the full Virasoro algebra, not just the global $sl(2,\mathbb{R})\times sl(2,\mathbb{R})$ subalgebra. Furthermore, the CFT can be seen to set some bulk properties naturally through boundary conditions. These properties provide an external source for choices that otherwise seem arbitrary. For example, we found no obvious physical difference between the Ishibashi and crosscap states, because we had the freedom to relabel algebra generators. In AdS, these generators have an independent physical meaning in the boundary CFT that must be matched, hence the statement that the point at the origin must be a crosscap state rather than an Ishibashi state.

We also performed an analytic continuation and considered singlet states in the Lorentzian case, where for illustration, we focused on the inflationary patch of Lorentzian dS$_3$. To describe gravity in Lorentzian de Sitter we were led to consider $SL(2,\mathbb{C})$ Chern-Simons theory. In this context, the choice of singlet state led to a natural reality condition for the $SL(2,\mathbb{C})$ Chern-Simons gauge fields. Lorentzian Wilson lines had a direct interpretation in terms of unitary $sl(2,\mathbb{R})$ representations that we motivated using an analytic continuation of our Euclidean $su(2)$ representations.  Since the inflationary patch has a large amount of apparent symmetry, it would also be interesting to repeat our analysis for less symmetric bulks such as Kerr-dS$_3$ \cite{Deser:1983nh}.

\paragraph{Bulk reconstruction in 3D de Sitter.}
The comparison to AdS/CFT naturally raises the question of bulk reconstruction. Consider our Lorentzian results for the inflationary patch. We now have an expression for pseudofields $|U(x)\rangle$ in terms of an abstract basis of states $|h,p,\bar{p}\rangle$ that mimics the discussion in AdS. And while a dS/CFT correspondence \cite{Strominger:2001pn,Witten:2001kn,Maldacena:2002vr} is far from established, suppose for the sake of argument that we take seriously the idea that our states $|h,p,\bar p\rangle$ can be described as operators in a putative CFT, in other words that there is a state-operator correspondence that maps our states to operators inserted at the origin $w,\bar{w}=0$: $|h,p,\bar p\rangle = \mathcal O(0,0)|0\rangle$. Then the Ishibashi state
\begin{align}
|\Sigma_{\rm Ish}\rangle &= \sum_{p=0}^\infty |h, p, p\rangle
\end{align}
can be expressed as
\begin{align}
|\Sigma_{\rm Ish}\rangle &= \sum_{p=0}^\infty \frac{\Gamma(2h)}{\Gamma(p+1) \Gamma(p+2h)}\mathcal H_{-1}^p \bar{\mathcal H}_{-1}^p \mathcal O(0,0)|0\rangle~. \label{eq:Ish-smeared}
\end{align}
On the other hand, the Ishibashi state can be thought of as being localized at a particular bulk point, as seen in (\ref{eq:ish-infl}). This suggests that we can obtain pseudofields at arbitrary bulk points by acting on both sides of (\ref{eq:Ish-smeared}) with $sl(2,\mathbb{R})$ generators. On the bulk side, this could be interpreted as diffeomorphisms that move the point while on the boundary side there is a natural interpretation in terms of conformal transformations.

Thus, we are led to ask: is there then an analogue of the HKLL procedure\cite{Hamilton:2005ju,Hamilton:2006az}, where local fields in de Sitter can be thought of as a smearing of states in a region of a lower-dimensional surface? And is there an implementation of that procedure in Chern-Simons theory?
To answer these questions, it is useful to compare to the existing literature on bulk reconstruction in de Sitter. A smearing function for the inflationary patch was constructed in~\cite{Xiao:2014uea}, and further developments include \cite{Sarkar:2014dma,Chatterjee:2016ifv,Anninos:2017eib}. Restricting to $d=2$, the result is that a local scalar field $\Phi$ of mass $m$ in  the inflationary patch of dS$_3$ can be represented as
\begin{align} 
\Phi(\eta, w, \bar w) = \int_{|w'\bar{w}'|<\eta^2} d w' d\bar{w}' &\Big[\frac{\Gamma(\Delta)}{\pi \Gamma(\Delta-1)}\left(\frac{\eta^2-|w' \bar{w}'|}{\eta}\right)^{\Delta-2} \mathcal O_+(w+ w', \bar{w}+ \bar{w}') \cr &+  \frac{\Gamma(2-\Delta)}{\pi \Gamma(1-\Delta)} \left(\frac{\eta^2-|w' \bar{w}'|}{\eta}\right)^{-\Delta} \mathcal O_-(w + w',\bar{w}+ \bar{w}') \Big]~.
\end{align}
In de Sitter it was crucial to keep the contributions from not only a scalar operator $\mathcal O_+$ with scaling dimension $\Delta_+ = \Delta = 1+ \sqrt{1-m^2\ell^2}$ dual to $\Phi$,  but also the shadow operator $\mathcal O_-$ with scaling dimension $\Delta_- = 2-\Delta= 1-\sqrt{1-m^2\ell^2}$. Here it is necessary to have these two contributions for the two-point function of the field to reproduce the correct Green's function, \eqref{eq:Greensfn}, which differs substantially from AdS. The difference is related to the fact that the Euclidean Green's function we use for de Sitter is not simply the analytic continuation of the AdS Green's function, which would violate microcausality~\cite{Bousso:2001mw}.

In our language the two terms come from considering the two representations with a fixed Casimir, with $l = -h$ and $l=h-1$. Other than this subtlety, and assuming the existence of a state operator correspondence for the states in our representations, the computation of the contribution to a bulk local field for each set of operators in terms of smearing functions proceeds exactly analogously to the Poincar\'{e} case considered in~\cite{Goto:2017olq}. All that is needed is to express the singlet state, translated to a point in the bulk, in terms of differential operators acting on CFT operators. This can then be converted into an integral representation in terms of smearing functions. There is however a need to have a more fundamental understanding of the role of $\mathcal O_+$ and $\mathcal O_-$ and its implications in dS quantum gravity.


\paragraph{Exact results in Chern-Simons theory.}
Chern-Simons theory on $S^3$, with a compact gauge group, is exactly solvable using the techniques of non-abelian localization~\cite{Beasley:2005vf}. In particular, the Wilson loop expectation value can be computed exactly in this context~\cite{Beasley:2009mb,Beasley:2010hm}. This suggests an extension of our semiclassical Euclidean results to a full quantum computation.  

There are two crucial differences in our approach that prevent us from applying exact results directly. The first is that we consider Wilson line operators rather than loops, which means that our probes are not gauge invariant. Additionally, we compute the Wilson line for infinite dimensional (and subsequently non-unitary) rather than finite dimensional representations of $su(2)$. The choice of this peculiar representation is in fact intricately linked to the non-gauge invariance of the Wilson lines, as we required infinite dimensional representations to construct the singlet states describing the endpoints. In the semiclassical version these limitations did not end up presenting an obstruction to a generalization as in \cite{Ammon:2013hba,Castro:2018srf}, and so it would be interesting to implement techniques of localization to construct and quantify our Wilson line as a quantum operator.

It would be especially interesting to see if the quantization of the Wilson line sheds light on the necessity in de Sitter of using two representations $\mathscr{R}_{\pm}$, which from the CFT standpoint led us to consider an additional set of shadow operators.  We saw that these were necessary in our framework to generate the complete set of quasinormal modes for de Sitter, and they are also crucial to reproduce the correct Green's function from a smearing function representation of a bulk local field. Moving beyond kinematics, one might hope that a quantization would help us define a Hilbert space that incorporates both representations and gives a definition for their overlap.

\section*{Acknowledgements}
We are  thank Costas Bachas, Monica Guica, Kurt Hinterbichler, Eva Llabr\'es,  and Alex Maloney for useful discussions. This work is supported by the Delta ITP consortium, a program of the Netherlands Organisation for Scientific Research (NWO) that is funded by the Dutch Ministry of Education, Culture and Science (OCW).  PSG thanks the IoP at the University of Amsterdam for its hospitality during this project and acknowledges the support  of  the  Natural  Sciences  and  Engineering  Research  Council  of  Canada (NSERC) [PDF-517202-2018]. AC thanks the Laboratoire de Physique Th\'eorique de l'Ecole Normale Sup\'erieure for their hospitality while this work was completed.

\appendix

\section{Conventions}\label{app:conv}

In this appendix we collect some basic conventions related to the Lie group $SU(2)$ and its algebra. 
For the algebra we  use generators $L_a$ and $\bar L_a$, $a=1,2,3$, and we have
\be \label{eq:su2xx}
[L_a, L_b] = i \epsilon_{abc} L^c~, 
\ee
with $\epsilon_{123}\equiv 1$. For the invariant bilinear form, we take
\be 
\mbox{Tr}(L_a L_b)  = {1\over 2}\delta_{ab}~.
\ee
Indices are  raised with $\delta^{a b}$. In the fundamental representation of $su(2)$, we have $L_a=\frac{1}{2}\sigma_a$ with the Pauli matrices given by
\begin{equation}
\sigma_1=\begin{bmatrix}
0 & 1\\
1 & 0
\end{bmatrix}~~,~~
\sigma_2=\begin{bmatrix}
0 & -i\\
i & 0
\end{bmatrix}~~,~~
\sigma_3=\begin{bmatrix}
1 & 0\\
0 & -1
\end{bmatrix}~~.
\end{equation}

To make an explicit distinction between the group and the algebra, we denote $G(M)$ as group element, and $L_a$ are the algebra generators as specified above. The general group action is given by
\be 
G(M^{-1}) L_a G(M) = D_a^{\ a'}(M) L_{a'}~, \indent \bar G(M^{-1}) \bar L_a \bar G(M) = D_a^{\ a'}(M) \bar L_{a'}~,
\ee
where $D's$ are the elements  in the adjoint representation of $su(2)$. As expected for any group, we also have
\be 
G(M_1)G(M_2) = G(M_1 M_2)~, \indent D_a^{\ b}(M_1)D_b^{\ c}(M_2) = D_a^{\ c}(M_1 M_2)~.
\ee

\section{Metric formulation of dS$_3$ gravity}\label{app:ds3}

\subsection{Coordinates and patches}\label{app:coord}
\begin{figure}
\begin{tabular}[c]{cc}
\begin{subfigure}[c]{0.5\textwidth}
\centering
\includegraphics[scale=0.66]{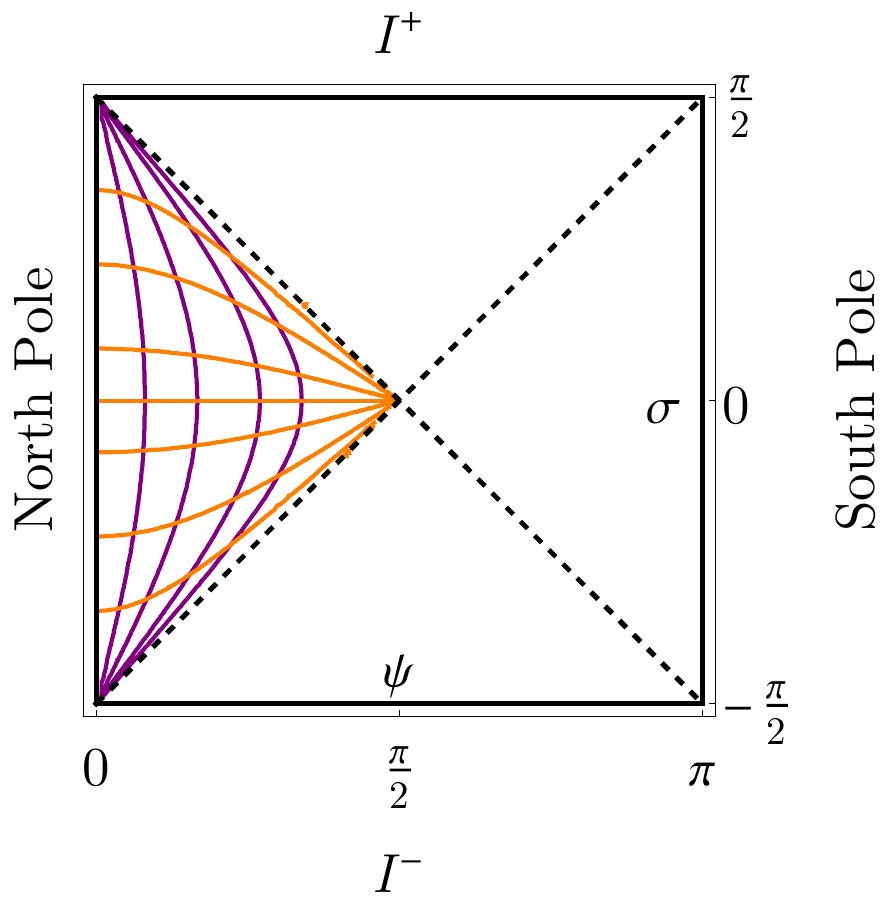}
\caption{static patch}
\label{fig:penrose_stat}
\end{subfigure} &
\begin{subfigure}[c]{0.5\textwidth}
\centering
\includegraphics[scale=0.66]{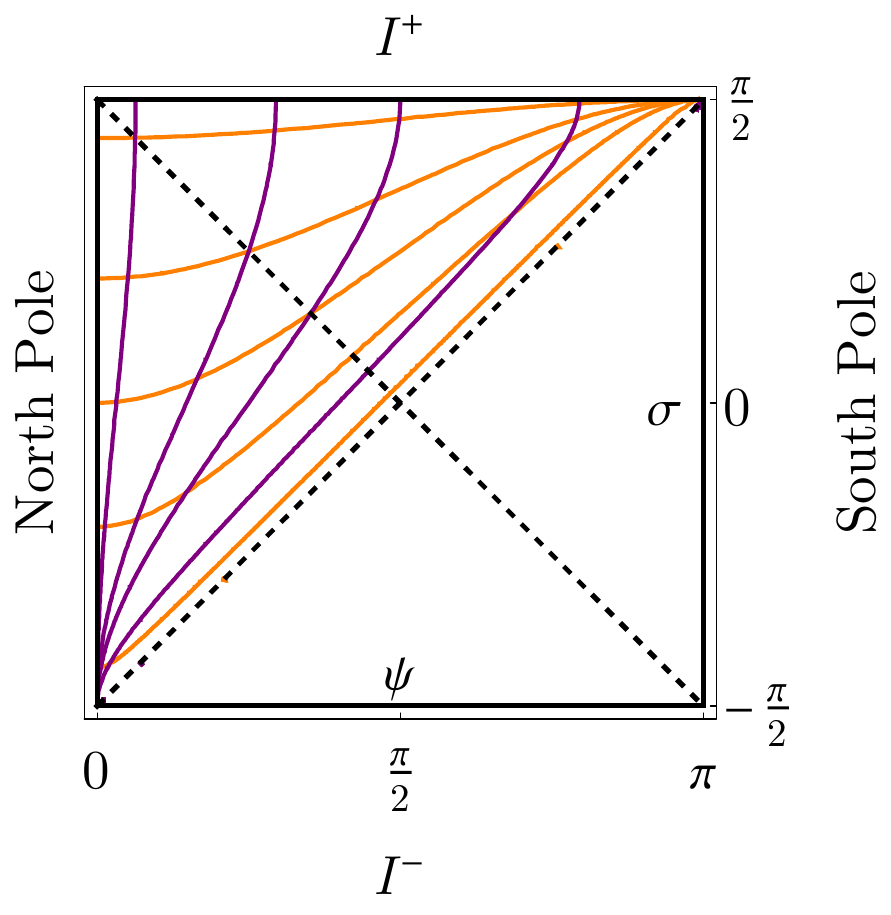}
\caption{inflationary patch}
\label{fig:penrose_infl}
\end{subfigure}
\end{tabular}
\caption{(Colour online) Penrose diagram of three-dimensional de Sitter space. Horizontal lines are slices of constant global time $T$ (or $\sigma$), which correspond to 2-spheres. $\psi$ is the polar angle on that sphere, so that each point on the diagram is a circle of radius $\sin \psi$. Vertical lines are slices of constant $\psi$. The top and bottom of the diagram are asymptotic timelike infinity, and the left and right edges are the North and South poles of the 2-spheres at each instant in global time. Constant $t$ (orange) and $r$ (or $u$, purple) slices on the static patch are shown on the static patch, with $r=0$ at the North Pole and $r$ increasing to $\frac{\pi}{2}$ at the horizon. Constant $\eta \geq 0$ (orange) and $x$ (for $y=0$, purple) slices are shown on the inflationary patch, with $\eta \rightarrow 0^+$ corresponding to positive timelike infinity and increasing to $+\infty$ at the horizon.}
\label{fig:penrose}
\end{figure}

Three-dimensional de Sitter is easily understood in terms of its embedding in four-dimensional Minkowski space:
\begin{equation}
-(X^0)^2+(X^1)^2+(X^2)^2+(X^3)^2=\ell^2~.
\end{equation}
Global dS$_3$ corresponds to the following parametrization, which covers the whole space-time:
\begin{align}
X^0&=\ell \sinh(T/\ell) ~,\cr
X^1&=\ell \cosh(T/\ell) \cos \psi~, \cr
X^2&=\ell \cosh(T/\ell) \sin \psi \cos \phi~,\cr
X^3&=\ell \cosh(T/\ell) \sin \psi \sin \phi~,
\end{align}
with $\psi$ and $\phi$ the polar and azimuthal coordinates of a two-sphere of unit radius. The metric is then
\begin{equation}
ds^2=-\dd T^2+\ell^2 \cosh^2(T/\ell) \left( \dd\psi^2 + \sin^2(\psi) \dd\phi^2 \right)~.
\end{equation}
The global time coordinate $T$, which has an infinite range, can be conformally rescaled:
\begin{equation}
\tan(\sigma)\equiv \sinh(T/\ell)~,~~ -\frac{\pi}{2}<\sigma <\frac{\pi}{2}~.
\end{equation}
After this rescaling, the metric is
\begin{equation}
ds^2=\frac{\ell^2}{\cos^2 \sigma}\left(-\dd\sigma^2+\dd\psi^2+\sin^2\psi \dd\phi^2\right)~.
\end{equation}
With the metric in this form, it is easy to draw the Penrose diagram in Fig.~\ref{fig:penrose}. 

Another useful parametrization of embedding coordinates is the following:
\begin{align}
X^0&=\sqrt{\ell^2-u^2}\sinh(t/\ell) ~,\nonumber\\
X^1&=\sqrt{\ell^2-u^2}\cosh(t/\ell) ~,\nonumber\\
X^2&=u \cos\phi ~,\nonumber\\
X^3&=u \sin \phi~,
\end{align}
for which the metric can be written as
\begin{equation}
ds^2=-\left(1-\frac{u^2}{\ell^2}\right)\dd t^2+\frac{\dd u^2}{1-\frac{u^2}{\ell^2}}+u^2\dd\phi^2~.
\end{equation}
This is the static patch of dS$_3$. It has the advantage of making a timelike Killing vector manifest, at the cost of covering only a portion of the whole manifold. We can see which portion by relating the two parametrizations:
\begin{align}
u&=\ell \cosh(T/\ell) |\sin(\psi)|=\ell \left|\frac{\sin(\psi)}{\cos(\sigma)}\right|~, \cr
\sinh^2(t/\ell)&=\frac{\sinh^2(T/\ell)}{1-\cosh^2(T/\ell)\sin^2(\psi)}=\frac{\sin^2(\sigma)}{\cos^2(\sigma)-\sin^2(\psi)}~.
\end{align}
In particular, for the embedding coordinates to be real, we need $0\leq u\leq \ell$. $u=0$ corresponds to $\psi=0$ and $u=\ell$ corresponds to $\sigma=\pm\left(\psi-\frac{\pi}{2}\right)$, so that these coordinates cover the left wedge of the Penrose diagram (or the right wedge, but not both if the coordinates are to be single-valued). Trajectories of constant $u$ or $t$ are shown in  Fig.\,\ref{fig:penrose_stat}. A simple coordinate redefinition brings us to the coordinates used in the main text:
\begin{equation}
u=\ell \sin(r)~.
\end{equation}
The embedding coordinates then take the form
\begin{align}
X^0&=\ell \cos(r) \sinh(t/\ell) ~,\nonumber\\
X^1&=\ell \cos(r) \cosh(t/\ell) ~,\nonumber\\
X^2&=\ell \sin(r) \cos(\phi) ~,\nonumber\\
X^3&=\ell \sin(r) \sin(\phi)~,
\end{align}
and the metric is
\begin{equation}
ds^2=-\cos^2(r) \dd t^2+\ell^2 \dd r^2+\ell^2 \sin^2(r) \dd\phi^2~.\label{eq:metricLor}
\end{equation}
It is instructive to go to Euclidean time in these coordinates: $t\rightarrow -i \tau \ell$,\footnote{Our Lorentzian metric has a mostly-$+$ signature. This fixes $t\rightarrow -i\tau \ell$ rather than $t\rightarrow +i\tau \ell$ in order to ensure that the equations of motion minimize the Hamiltonian rather than maximize it. The factor of $\ell$ is there to make the interpretation of $\tau$ as an angular coordinate manifest.} which leads to
\begin{equation}
{ds^2\over \ell^2}=\cos^2(r) \dd \tau^2+ \dd r^2+  \sin^2(r)\dd\phi^2~,
\end{equation}
and
\begin{align}
X^1&=\ell \cos(r) \cos(\tau)~, \nonumber \\
X^2&=\ell \sin(r) \cos(\phi)~, \nonumber\\
X^3&=\ell \sin(r) \sin(\phi)~, \nonumber\\
X^4&=\ell \cos(r) \sin(\tau)~ , \label{eq:hopf-coords}
\end{align}
where we've defined $X^4=iX^0$. These coordinates are simply the Hopf coordinates for a three-sphere embedded in $\mathbb{R}^4$. Avoiding a conical singularity near $r=\frac{\pi}{2}$ requires that $\tau\sim \tau+2\pi$, from which we can read off the inverse temperature of the horizon: $\beta=2\pi \ell$.

Another parametrization of dS$_3$ gives coordinates on the inflationary patch:
\begin{align}
X^0&=-\ell \frac{\eta^2-1-x^2-y^2}{2\eta} \nonumber \\
X^1&=\ell \frac{\eta^2+1-x^2-y^2}{2\eta} \nonumber  \\
X^2&=\ell \frac{x}{\eta} \nonumber \\
X^3&=\ell \frac{y}{\eta} ~. \label{eq:infl-coords}
\end{align}
The metric in these coordinates is
\begin{equation}
\frac{ds^2}{\ell^2}=\frac{-d\eta^2+dx^2+dy^2}{\eta^2}~.
\end{equation}
With $0<\eta<\infty$, these coordinates cover half of the space-time, with $\eta^{-1}=\frac{\cos \psi}{|\cos \sigma|}+\tan\sigma$. $\eta=0^+$ corresponds to $\sigma=\frac{\pi}{2}$ (i.e. positive timelike infinity) and $\eta \rightarrow +\infty$ to $\sigma \rightarrow (\psi-\frac{\pi}{2})^+$. This is shown in Fig. \ref{fig:penrose_infl}.

\subsection{Geodesics and Green's functions in dS$_3$}
\label{sect:prop}
We now write down the propagator for a scalar field in the static patch of three-dimensional de Sitter. We can exploit the symmetry of the system to write the wave equation in terms of the geodesic distance between two points. This is easier to do in Euclidean signature, where we consider $S^3$ described by embedding coordinates $X^i$ given by  equation (\ref{eq:hopf-coords}).
The only invariant quantity we can write out of two vectors $X^i$ and $Y^i$ is $X \cdot Y$. In fact, the geodesic distance between two points is simply
\begin{equation}
\ell \Theta= \ell \arccos \left(\frac{X \cdot Y}{\ell^2}\right)~.
\end{equation}

The Euclidean propagator obeys:
\begin{equation}
\nabla^2 G(X,Y) - m^2 G(X,Y) = \delta^{(3)}(X-Y)~.
\end{equation}
The propagator can only depend on coordinates through the quantity $\chi=\cos(\Theta)$. This implies
\begin{align}
\nabla^2 G(X,Y)&= \nabla^2 G(\chi)\cr
&=\frac{1}{\sin ^2(\Theta)} \frac{d}{d\Theta} \left[(\sin(\Theta))^2 \frac{d G(\chi)}{d\Theta}\right] \cr
&=(1-\chi^2)\frac{d^2G}{d\chi^2} - 3\chi \frac{dG}{d\chi}~.
\end{align}
Therefore, the homogeneous version of the wave equation is
\begin{equation}
(1-\chi^2)\frac{d^2G}{d\chi^2} - 3\chi \frac{dG}{d\chi}-(m\ell)^2 G(\chi)=0~.
\end{equation}
This has solutions of the  following form:
\begin{equation}
G(\chi)=c_1 \frac{P\left(\sqrt{1-(m\ell)^2}-\frac{1}{2},\frac{1}{2};\chi\right)}{(\chi^2-1)^{1/4}} + c_2 \frac{Q\left(\sqrt{1-(m\ell)^2}-\frac{1}{2},\frac{1}{2};\chi\right)}{(\chi^2-1)^{1/4}}~,
\end{equation}
where $P$ and $Q$ are associated Legendre polynomials.
These associated Legendre polynomials simplify precisely when the second argument is $1/2$: 
\begin{align}
P(x,1/2,\cos(\Theta))&=\sqrt{\frac{2}{\pi \sin(\Theta)}} \cos\left(\left(x+\frac{1}{2}\right)\Theta\right)~,\cr
Q(x,1/2,\cos(\Theta))&=-\sqrt{\frac{\pi}{2 \sin(\Theta)}} \sin\left(\left(x+\frac{1}{2}\right)\Theta\right)~.
\end{align}
Therefore, introducing two new undetermined constants, we have
\begin{equation}
G(\Theta)=\frac{A}{\sin(\Theta)}\left[\cos\left(\left(\sqrt{1-(m\ell)^2}\right) \Theta\right) - C\sin\left(\left(\sqrt{1-(m\ell)^2}\right)\Theta\right)\right]~. \label{eq:g_gen}
\end{equation}

Short distances correspond to $\chi$ approaching $1$ from below, in other words taking $\Theta\sim \epsilon$. In that regime, we get
\begin{equation}
G(\Theta=\epsilon)=\frac{A}{\epsilon}+\cdots~.
\end{equation}
In three dimensions, we know that a properly normalized Green's function has a short-distance divergence that goes as $-\frac{1}{4\pi d}=-\frac{1}{4\pi \ell \epsilon}$, so this fixes
\begin{equation}
A=-\frac{1}{4\pi \ell}~.
\end{equation}
Therefore, the Euclidean propagator is
\begin{equation}
G(\Theta)=\frac{i}{4\pi \ell} \left[ (1+iC)\frac{e^{2ih\Theta}}{1-e^{2i\Theta}} - (1-iC) \frac{e^{-2ih\Theta}}{1-e^{-2i\Theta}}\right]~,
\end{equation}
with
\begin{equation} \label{eq:defnh}
h=\frac{1+\sqrt{1-(m\ell)^2}}{2}~.
\end{equation}
Choosing the value of the integration constant $C$ corresponds to picking a particular vacuum. The natural choice is $C=\cot(2h\pi)$. This removes the singularity at $\Theta=\pi$, in other words the singularity that appears on the lightcone of antipodal points. It is convenient to define
\begin{equation}
\mathcal{G}_h(\Theta)=\frac{i}{2\pi \ell} \frac{1}{1-e^{-4\pi i h}} \frac{e^{-2ih\Theta}}{e^{-2i\Theta}-1}~.
\end{equation}
The Green's function satisfying the boundary conditions we've outlined above naturally splits into
\begin{align}
G(\theta)&=\mathcal{G}_h(\Theta)+\mathcal{G}_{1-h}(\Theta) \\
&=-\frac{ \csc{(2\pi h)} \csc( \Theta) \sin\left(2h(\pi-\Theta)+\Theta\right)}{4\pi \ell}~ \\
&=-\frac{\Gamma(2h) \Gamma(2-2h)}{(4\pi)^{3/2} \Gamma(3/2)} {~}_2F_1\left(2h,2-2h ; \frac{3}{2},\frac{1+\cos(\Theta)}{2}\right)~,\label{eq:Greensfn}
\end{align}
where the last line is manifestly the Green's function in the Euclidean vacuum of dS$_3$\cite{Bousso:2001mw}.

We can write $\Theta$ explicitly in terms of the Hopf coordinates (\ref{eq:hopf-coords}):
\begin{equation}
\cos(\Theta)= \cos(r_i)\cos(r_f)\cos(\Delta \tau)+\sin(r_i)\sin(r_f)\cos(\Delta \phi)~.\label{eq:distace} 
\end{equation}
Finally, note that if we want to analytically continue back to Lorentzian signature, we should take $\tau \rightarrow i t/\ell$. In that case, we notice that $|\cos(\Theta)|>1$ whenever points are timelike separated, in which case $i \ell \Theta$ corresponds to the proper time between the points. When the points are spacelike separated, $\ell \Theta$ remains the proper length between them. 

In terms of inflationary coordinates (\ref{eq:infl-coords}), the invariant distance is
\begin{align}
\cos(\Theta)&=\frac{\eta_i^2+\eta_f^2-(\Delta x)^2-(\Delta y)^2}{2\eta_i \eta_f} \\
&=1+ \frac{(\Delta \eta)^2-(\Delta x)^2-(\Delta y)^2}{2\eta_i \eta_f}~,
\end{align}
where the distinction between spacelike-separated and timelike-separated points is manifest.

\section{Analytic continuation in the Chern-Simons formulation}\label{app:cont}
Here, we provide more details on how to construct an analytic continuation between Euclidean and Lorentzian signature from the Chern-Simons perspective. The analytic continuation from Euclidean to Lorentzian signature is most easily understood in terms of these generators, which are simply related to rotations and boosts in embedding space.

The Euclidean Chern-Simons action, \eqref{eq:CSaction} and~\eqref{eq:SCS}, can be written in unsplit form as
\be S_{E} = -\frac{k}{4\pi} \int \mbox{Tr}\left(\mathscr{A} \wedge d\mathscr{A} + \frac{2}{3} \mathscr{A}\wedge \mathscr{A} \wedge \mathscr{A}\right)~,\label{eq:CSactionEucl}\ee
where the gauge field is expanded in terms of the generators of Euclidean $so(4)$ isometries as
\be \mathscr{A} = e^a P_a + \omega^a J_a~,\ee
with
\begin{align}
[J_a, J_b] &= -\epsilon_{abc} J^c~,\\
[J_a, P_b] &= -\epsilon_{abc} P^c~,\\
[P_a,P_b] &= -\Lambda \epsilon_{abc} J^c~,
\end{align} 
and indices are raised with $\delta^{ab}$. To construct an analytic continuation, we need to find a map to generators of the $so(1,3)$ algebra of isometries of Lorentzian dS$_3$,
\begin{align}
[\mathcal J_a, \mathcal J_b] &= \epsilon_{abc} \mathcal J^c~,\label{eq:so311}\\
[\mathcal J_a, \mathcal P_b] &= \epsilon_{abc} \mathcal P^c~,\\
[\mathcal P_a,\mathcal P_b] &= -\Lambda \epsilon_{abc} \mathcal J^c~,\label{eq:so313}
\end{align}
with indices raised by $\eta^{ab}$. 

One possibility is given by the following:
\begin{align}
\mathcal J_1 &= i P_1~,\nonumber \\
\mathcal J_2 &= i J_2~, \nonumber \\
\mathcal J_3 &= -P_3~, \nonumber\\
\mathcal P_1 &=  J_1~, \nonumber\\
\mathcal P_2 &= P_2~, \nonumber\\
\mathcal P_3 &= i J_3~.
\end{align}
Under this map, the $SO(4)$ bilinear form
\begin{align} 
\ell \,\mbox{Tr}(J_a P_b) = -\delta_{ab}~, \indent \mbox{Tr}(J_a J_b) = \ell^2\,\mbox{Tr}(P_a P_b) = 0~\label{eq:bilinearEucl}
\end{align}
gets taken to 
\begin{align} 
\ell \,\mbox{Tr}(\mathcal J_a \mathcal P_b) =  - i \, \eta_{ab}~, \indent \mbox{Tr}(\mathcal J_a \mathcal J_b) = \ell^2\,\mbox{Tr}(\mathcal P_a \mathcal P_b) = 0~.\label{eq:bilinearLor}
\end{align}

While the map we have constructed can be viewed as a map between real algebras, \eqref{eq:bilinearLor} is not an invariant bilinear form for real $SO(3,1)$. Indeed, the unique invariant bilinear form for $SO(3,1)$ is given by
\be \left<\mathcal J_a, \mathcal J_b\right>=\eta_{ab}~, \indent \left<\mathcal P_a, \mathcal P_b\right> = -\Lambda \eta_{ab}~, \indent \left<\mathcal J_a, \mathcal P_b\right> = 0~,\label{eq:bilinear2}\ee
rather than \eqref{eq:bilinearLor}. In the Chern-Simons formulation for gravity one typically chooses a $\mbox{Tr}(J_a P_a)$ bilinear form for a reason, as the Chern-Simons theory defined using \eqref{eq:bilinear2} does not reduce to Einstein gravity (see~\cite{Witten:1988hc}). It is for this reason that we have considered a complexification to $SL(2,\mathbb{C})$. While the real $SO(3,1)$ algebra does not split as in \eqref{eq:LLbar}, the complexification does split and therefore admits multiple bilinear forms, not only \eqref{eq:bilinear2} but also \eqref{eq:bilinearLor}. 

With the map defined above, the bilinear form for the barred sector has the wrong sign:
\be \mbox{Tr}(\bar{\mathcal L}_a \bar{\mathcal L}_b) = +\frac{1}{2} \eta_{ab}~. \ee
We can flip the sign while simultaneously multiplying the barred action by a minus sign. Combined with an analytic continuation of the Chern-Simons coupling,
\be s=ik~,\ee
this takes us from \eqref{eq:CSactionEucl} to the $SL(2,\mathbb{C})$ Chern-Simons action, \eqref{eq:CSactionLor}.

\bibliographystyle{JHEP-2}
\bibliography{ref}

\end{document}